\documentclass[%
 aip,
 amsmath,amssymb,
 reprint,%
]{revtex4-1}

\usepackage{graphicx}% Include figure files
\usepackage{dcolumn}% Align table columns on decimal point
\usepackage{bm}% bold math
\usepackage[utf8]{inputenc}
\usepackage[T1]{fontenc}
\usepackage{mathptmx}
\usepackage{etoolbox}

\usepackage{placeins}
\usepackage{amsmath}
\usepackage{mathtools}
\makeatletter
\def\@email#1#2{%
 \endgroup
 \patchcmd{\titleblock@produce}
  {\frontmatter@RRAPformat}
  {\frontmatter@RRAPformat{\produce@RRAP{*#1\href{mailto:#2}{#2}}}\frontmatter@RRAPformat}
  {}{}
}%
\makeatother
\begin{document}

\preprint{AIP/123-QED}
% Numerical modelling of magnetohydrodynamically affected shock interactions in hypersonic flows 
% Numerical modelling of imposed magnetohydrodynamic effects on shock interactions in hypersonic flows 
%Numerical modelling of shock interactions in hypersonic flows with magnetohydrodynamic effects
\title[Numerical modelling of imposed magnetohydrodynamic effects in hypersonic flows]{Numerical modelling of imposed magnetohydrodynamic effects in hypersonic flows}  
% Force line breaks with \\

\author{ \vspace{2mm} H. A. Muir}
 \email{ham38@cam.ac.uk}
 \affiliation{Cavendish Laboratory, Department of Physics, University of Cambridge}%Lines break automatically or can be forced with \\
\author{N. Nikiforakis}%
%\altaffiliation[Also at ]{Cavendish Laboratory, Department of Physics, University of Cambridge}

\date{\today}% It is always \today, today,

\begin{abstract}
Weakly ionised plasmas, formed in high enthalpy hypersonic flows, can be actively manipulated via imposed magnetic fields - a concept termed magnetohydrodynamic (MHD) flow control. Imposed MHD effects, within flows which exhibit multiple shock interactions, are consequential for emerging aerospace technologies: including the possibility of replacing mechanical control surfaces with \emph{magnetic actuation}. However, numerical modelling of this flow type remains challenging due to the sensitivity of feature formation and the real gas modelling of weakly ionised, electrically conductive, air plasma. 
In this work, numerical simulation capabilities have been developed for the study of MHD affected, hypersonic flows, around 2D axisymmetric non-simple geometries. The validated numerical methodology, combined with an advanced 19 species equation of state for air plasma, permits the realistic and efficient simulation of air plasmas in the equilibrium regime. Quantitative agreement is achieved between simulation and experiment for a Mach 5.6 double cone geometry with applied magnetic field. 
In the context of the magnetic actuation concept, numerical studies are conducted for varied conical surface angle and magnetic field configuration. %The studies reveal detailed, and at times counter-intuitive, flow field effects. 
For simple geometries with an elemental shock type, the MHD enhancement effect produces a self-similar shock structure. This paper demonstrates how, for hypersonic flows with complex shock interactions, the MHD affected flow is not only augmented in terms of shock position, but may exhibit topological adaptations in the fundamental flow structure. A classification system is introduced for the emergent flow topologies identified in this work. Fluid-magnetic interactions are explored and explained in terms of the coupled mechanisms leading to: (1) differences in magnitude of MHD enhancement effect, and (2) structural adaptations of the flow topology. The applied numerical studies examine: why increased conical surface angle does not amplify the MHD enhancement effect as expected from the base flow conditions, and the mechanisms by which the magnetic field configuration influences the MHD augmented shock structure. Most critically, classes of conditions are identified which produce topological equivalence between the magnetic interaction effects and a generalised mechanical control surface. 
\end{abstract}

\maketitle

\section{Introduction}
\label{sec:Introduction} 

In high enthalpy hypersonic flows, the pressures and temperatures in the shock layer can become sufficiently high that dissociation and ionisation reactions produce a weakly ionised plasma. Owing to its elevated electrical conductivity, the plasma can be actively manipulated via electromagnetic effects when a strong magnetic field is generated from within the vehicle. 

\begin{figure}[h!]
	%\vspace{-2mm}
	\centering
	\includegraphics[width=7.7cm]{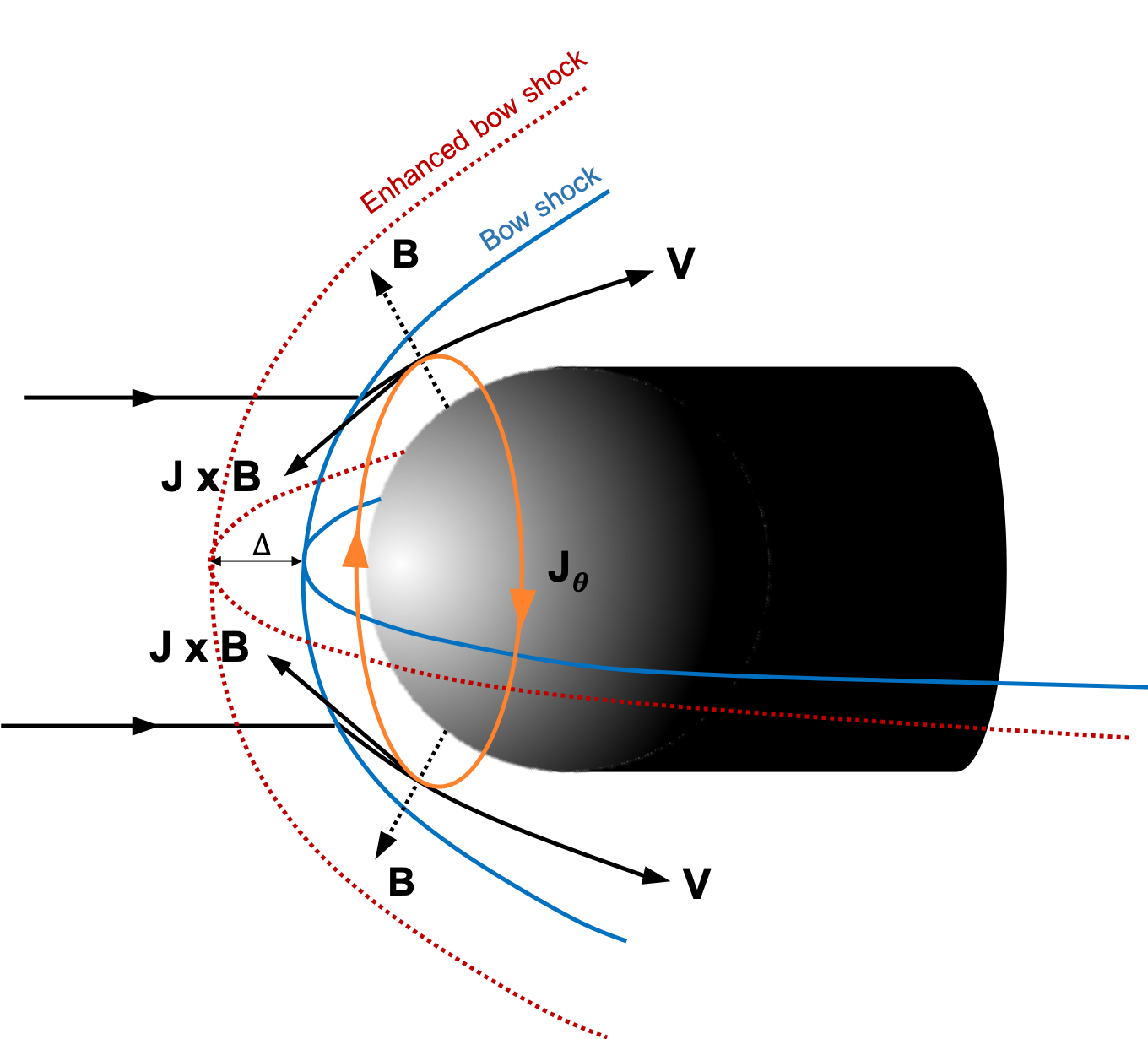}
	\vspace{-2mm}
	\caption{MHD flow control concept: streamlines (\textbf{V}) are shown, carrying charged particles inside the bow shock. Two local tangential components of the larger dipole magnetic field are labelled \textbf{B}. The induced circumferential electric current $\textbf{J}_\theta$ crosses through the magnetic field to produce a Lorentz force in the $\textbf{J}\times\textbf{B}$ direction, enhancing the shock stand-off distance by $\Delta$. }
	\label{fig:MHDcontrol_diagram}
	\vspace{-3mm}
\end{figure}

The MHD flow control concept is depicted in FIG. \ref{fig:MHDcontrol_diagram} for a generic blunt body. The charged particles travelling within the shock layer pass through the magnetic field lines (\textbf{B}), which induces a strong circumferential electric current (\textbf{J}). This electric current interacts with the magnetic field to produce a Lorentz force ($\textbf{J} \times \textbf{B}$) which acts macroscopically on the fluid. The resultant effect is to force the bow shock further away ($\Delta$) from the vehicle surface (increasing the bow shock total surface area), and is termed \emph{shock layer enhancement} \cite{Matsuda2006}. 

\begin{figure*}[ht]
	\centering
	\includegraphics[width=14.0cm]{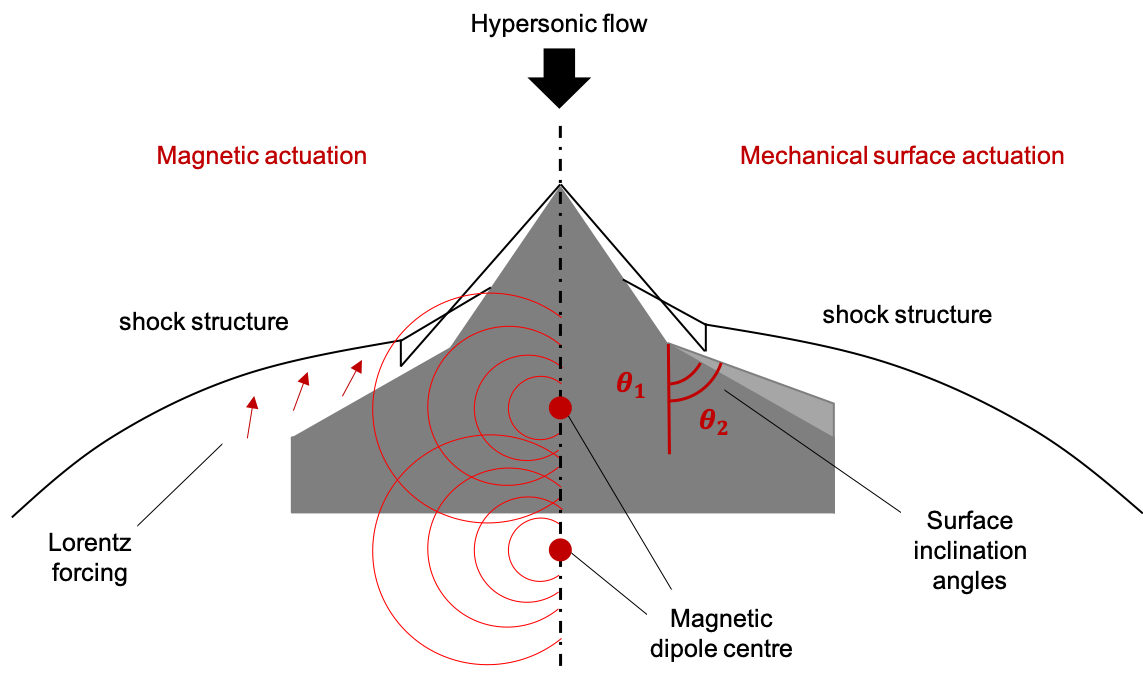}
	\caption{Comparison of magnetic actuation via MHD effects, and mechanical actuation via physical surface inclination, for a double cone geometry. Left) two different positions of magnetic dipole centre are shown, where resultant Lorentz forcing acts on the shock layer, and right) two different second surface inclination angles are depicted ($\theta_1$ and $\theta_2$), which similarly influences the shock position. } 
	\label{fig:surface_actuation_diagram}
\end{figure*}

At present, prospective applications of MHD flow control for aerospace technologies are an active area of research. These include: the MHD by-pass engine (acts to accelerate ionised flow out of a combustor to generate additional thrust)  \cite{Kaminaga2005}$^,$\cite{Park2003bypass}, hypersonic inlet control for scramjet engines (the leading oblique shock angle is magnetically controlled to achieve on-design specification at off-design flight condition) \cite{Sheikin2004, Shneider2003, Macheret2004}
%Kuranov2003,$^,$\cite{Bityurin2003}$^,$\cite{Kuranov2003}$^,$\cite{Sheikin2004}$^,$\cite{Shneider2003}$^,$\cite{Shneider2004}$^,$\cite{Macheret2004},Matsuda2008, 
and the prospect of active thermal protection via magnetic heat shielding \cite{Yoshino2009, Otsu2010, Li2017} as well as drag enhancement \cite{Kawamura2009, Shimosawa2016} for atmospheric reentry vehicles. In flows with insufficient ionisation, alternative methods are used to locally elevate electrical conductivity. Such methods (including plasm actuators in the boundary layer) often generate an additional electric field, which creates a supplementary mechanism in the Lorentz force \cite{Shang2005}. In hypersonic flows where a sufficiently high level of electrical conductivity is generated via direct impact of the freestream with the vehicle body, the imposed magnetic field can directly affect the flow \cite{MHD_report} - this is the focus of this work. 

%In terms of generic aerodynamic control surfaces, experimental and numerical studies have examined the MHD control of the leading oblique shock over various wedge geometries \cite{Bityurin1997, Bityurin2004}, and conical bodies \cite{Bobashev2006, Baccarella2011}. 
%MHD control of the boundary layer in hypersonic flows tend to utilise \emph{plasma actuators}, where electrodes are placed on the surface to ionise the boundary layer and create an added mechanism in the Lorentz force due to the generated electric field \cite{Shang2005}.  
%Such methods are found to be viable and power efficient \cite{Gong2020, Borghi2003} - low input energy is required due to the low fluid momentum in the near-surface boundary layer (reduced required Lorentz force to affect the flow). 
%Plasma actuators have been applied control shock wave boundary layer (SWBL) interactions which occur in hypersonic inlets where an oblique shock wave impinges upon a secondary surface \cite{Deshpande2018, Jiang2020}. 
%Due to the cost and difficulty of real flight and experimental research, computational models play an important role in the study of prospective MHD applications. 

Extreme forces are experienced along the surfaces of hypersonic vehicles. Control flaps and surfaces require high power mechanical actuators and multiple moving parts to adjust surface angles dynamically during flight. MHD flow control presents the desirable possibility of replacing mechanical actuation of the control surface, however, such research is in the early stages\cite{Wasai2010}. 

Experimental and numerical research for MHD control of hypersonic flows has generally focussed on simple geometries \cite{Baccarella2011} which produce an elementary shock structure: such as the single bow shock formed around blunt bodies\cite{Coakley1971, Poggie2002} and the leading oblique shock formed over wedges\cite{Bityurin2004, Lineberry2004} and cones\cite{Bobashev2006, Baccarella2011}. In the case of multiple angled surfaces - which realistically represents a mechanical control surface, such as a flap or ramp - the coupled fluid dynamic and magneto-dynamic effects become complex. Shock wave boundary layer (SWBL) interactions lead to the formation of a separation region, and produces multiple shock interactions. With sufficient ionisation in the flow, the high conductivity low velocity regions which form within the steady state shock structure, are sensitive to manipulation via magnetic fields \cite{Shang2018}. 

The double cone is considered a useful test geometry since the resultant flow exhibits all of the relevant features of common 3D internal engine geometries or aerodynamic control surfaces. FIG. \ref{fig:surface_actuation_diagram} depicts the comparison between shock structure control via imposed MHD effects and traditional mechanical surface actuation.

%These are considered regions of high \emph{magnetic interaction}. 
%Collectively, there is extensive numerical research on MHD flow control for blunt body reentry vehicles, leading oblique shock control for scramjet engine intakes, and MHD surface actuation of boundary layers within hypersonic flows. In contrast, there is a paucity of research which deals with MHD control of flows with shock-shock and shock wave boundary layer interactions, where weakly ionised plasma is formed in multiple high temperature regions in the flow. Such interactions occur in complex flows around non-simple geometries such as double cones, and multiple angled surfaces, which are common in the design of hypersonic vehicles, and is a more realistic configuration for an aerodynamic control surface. 

%\begin{figure}[h!]
%	\centering
%	\includegraphics[width=8.5cm]{./vis/MHD_scramjet.png}
%	\caption{MHD assisted oblique shock control of scramjet inlet: left configuration shows ingested leading oblique shock when the system dynamic pressure is not precisely the on-design specification. This can be dynamically corrected via application of an imposed magnetic field generated within the vehicle.}
%	\label{fig:MHD_scramjet}
%\end{figure}

In the case without magnetic fields, detailed experimental research has been conducted on the 2D axisymmetric double cone \cite{CUBRC2014}. Even without electromagnetic interactions, high enthalpy laminar flows with SWBLI and real gas effects are challenging to numerically simulate, and have been the subject of many studies \cite{Hao2017, Zuo2021, Holloway2020}. 

In the case with a magnetic field generated from within the model, experiments were conducted by Wasai et al. \cite{Wasai2010} in an advanced expansion tube facility, for a hypersonic flow (with air as the test gas) over a double cone geometry. The experiments were able to demonstrate and measure an MHD enhancement effect due to the applied magnetic field on the flow which exhibits multiple shock interactions. Simulations of the experimental conditions were conducted by Nagata et al. \cite{Nagata2011numerical, Nagata2013Bfield} who were able to demonstrate similar qualitative effects as observed in the experiment. 
%Quantitative agreement between simulation and experiment, however, remains illusive. 

A number of prevailing numerical modelling challenges have been identified for this flow type. These include: the balance between computational tractability and physical accuracy of the governing mathematical model (the coupled fluid-electromagnetic flow physics exhibits complex multiscale phenomena), realistic thermochemistry modelling for weakly ionised air plasmas, accurate resolution of electric transport properties (which is typically related to the thermochemistry model), and numerical solvers which are robust for the hypersonic regime over complex geometries, including the stable implementation of Lorentz forcing dynamics. 

In the current work, 2D-axisymmetric simulations are conducted to investigate the physics of hypersonic flows over double cone geometries, with imposed MHD effects. Notably, the developed model includes a new extended 19 species equation of state (EoS) for air plasma applied to hypersonically formed plasmas in the equilibrium regime. This EoS enables the direct and accurate computation of electrical conductivity which is a key parameter driving the Lorentz forcing dynamics. The numerical solvers are implemented within the AMReX framework \cite{AMReX2019}, which permits block-structured parallelism and hierarchical adaptive mesh refinement, resulting in a high effective resolution at low cost and scalability of the computation. Efficient and robust embedded boundary geometry (via a rigid body Ghost Fluid Method) permits arbitrarily complex geometries to be simulated within the structured Cartesian mesh, and are shown to be stable for strong shock-body interactions. 

The numerical model is validated against experimental results. The model is then applied to study the double cone configuration with imposed MHD effects. Previous MHD studies on simple geometries - blunt bodies, single cones and wedges - produce an elemental shock structure: bow shock or oblique shock. When the magnetic field is imposed upon the flow, this causes a self-similar MHD enhancement effect: congruent shock structure, but with a change of position. In these cases, the enhancement can be effectively quantified through the measured change in one variable: \% increase in shock stand-off distance along the stagnation line for a blunt body, or the increase in angle of an oblique shock wave for conical or wedge geometries. The numerical studies of this work, however, demonstrate how, for a complex flow field combined with an imposed magnetic field, the base flow and the MHD affected flow can become structurally dimorphic. That is, the topology of the magnetically augmented flow is often \emph{not} self-similar to its respective base flow case. 

Formational differences in the flow structure are critical within a context of aerodynamic prediction for flight control. Therefore, this work examines the flow physics of the complex, and coupled, magneto-fluid behaviours with a focus on the conditions and underlying mechanisms which lead to structural variation of the steady state flow fields. In terms of the magnetic actuation control surface concept, not only the measurement of MHD shock enhancement effect, but the identification of structurally similar flow morphologies between MHD actuation and mechanical surface actuation has technological significance. 

%\FloatBarrier

\section{Mathematical model}

The magnetohydrodynamic behaviour in a compressible flow regime exhibits interdependent phenomena across disparate scales. For numerical solution to be tractable, assumptions and simplifications are necessary. \par 

In the case of aerospace plasmas - where collision frequency within the plasma is sufficiently high, flow length scales are significantly larger than the Debye length, and time scales are larger than the reciprocal of the plasma resonant frequency  - a continuum magnetohydrodynamic (MHD) model governs \cite{Goedbloed}. \par

Broadly across the literature, hypersonic plasmas are considered in the low magnetic Reynolds number regime \cite{MHD_report, Dias_MHD, Fujino2006} where $Re_m$ is of the order of unity or less. Under this regime the imposed and induced magnetic fields are sufficiently decoupled for the static imposed magnetic field to be unaffected by the flow. \par 

\vspace{-3mm}
\begin{equation}
Re_m = \frac{\text{Magnetic induction}}{\text{Magnetic diffusion}} = \frac{\sigma_0 V_0 L_0}{\epsilon_0 c^2} 
\end{equation}
\vspace{-3mm}

Where $\sigma_0$ is the nominal fluid electrical conductivity, $V_0$ the total velocity, $L_0$ the characteristic length, $\epsilon_0$ is the air permittivity, and $c$ the speed of light in a vacuum.  

The governing system equations are derived by combining the Navier-Stokes equations for fluid dynamics (accounting for fluid viscosity and thermal conductivity) with Maxwell's equations for electromagnetism. The derivation under the low $Re_m$ weakly ionised plasma regime results in resistive MHD equations which account for Lorentz forcing and Joule heating. 

An important dimensionless number throughout this work is the magnetic interaction parameter ($Q_{MHD}$) defined as:
\par

\vspace{-3mm}
\begin{equation}
Q_{MHD} = \frac{\text{Lorentz force}}{\text{Fluid inertial force}} = \frac{\sigma_0 B_0^2 L_0}{\rho_\infty V_\infty} 
\label{eq:QMHD}
\end{equation}
\vspace{-3mm}

Where $B_0$ is the peak magnetic field strength. The ability of the imposed magnetically field to dynamically affect the flow is best determined by $Q_{MHD}$.

\subsection{Governing system equations}

For studies of axisymmetric vehicle geometries, the axis of symmetry can be exploited, converting the 3D problem into a 2D plane in cylindrical coordinates. Therefore the MHD system equations are given in r-z space as: \par

\begin{equation}
\frac{\partial \textbf{U} }{\partial t} + \frac{\partial \textbf{F}}{\partial r} + \frac{\partial \textbf{G}}{\partial z} = \frac{\partial \textbf{F}_v}{\partial r} + \frac{\partial \textbf{G}_v}{\partial z} + \textbf{S}^c + \textbf{S}^c_v +\textbf{S}_{MHD}
\label{eq:governing_system}
\end{equation}

\textbf{U}, \textbf{F} and \textbf{G} denote the state vector, and r-z directional inviscid flux vectors respectively, and $\textbf{S}^c$ denotes the associated cylindrical source term vector. Similarly,  $\textbf{F}_v$ and $\textbf{G}_v$ denote  the r-z directional viscous flux vectors and associated cylindrical source term $\textbf{S}^c_v$. $\textbf{S}_{MHD}$ denotes the MHD source term vector. 

For the 2D r-z system, we denote the velocity vector in the $(\hat{r}, \hat{\theta}, \hat{z})$ direction to have components $(u,w,v)$, and so the system vectors are defined as: 

\def\Uc{
	\begin{pmatrix}
		\rho \\
		\rho u \\
		\rho w \\
		\rho v \\
		E \\
	\end{pmatrix}
}

\def\Fr{
	\begin{pmatrix}
		\rho u \\
		\rho u^2 + p \\
		\rho u w \\
		\rho u v \\
		u_r(E+p) \\
	\end{pmatrix}
}

\def\Gz{
	\begin{pmatrix}
		\rho w \\
		\rho u v\\
		\rho v w \\ 
		\rho v^2 + p \\
		v(E+p) \\
	\end{pmatrix}
}

\def\ScE{
	\begin{pmatrix}
		- \frac{1}{r} \rho u \\
		- \frac{1}{r} \rho u^2 \\
		- \frac{1}{r} \rho u w \\
		- \frac{1}{r} \rho u v \\
		- \frac{1}{r} u (E+p) \\
	\end{pmatrix}
}

\def\Fvt{
	\begin{pmatrix}
		0 \\ 
		\tau_{rr} \\
		\tau_{rz} \\
		\tau_{r\theta} \\
		Q_{r}  %u \tau_{rr} + v \tau_{rz} + w \tau_{r\theta} + q_r
\end{pmatrix}}

\def\Fv{
	\begin{pmatrix}
		0 \\ 
		\mu (\frac{4}{3} u_r - \frac{2}{3} v_z - \frac{2}{3} \frac{u}{r})\\
		\mu (v_r +  u_z)\\
		\mu (w_r - \frac{w}{r})\\
		u \tau_{rr} + v \tau_{rz} + w \tau_{r\theta} + \zeta T_r
\end{pmatrix}}

\def\Gvt{
	\begin{pmatrix}
		0 \\ 
		\tau_{zr} \\
		\tau_{zz} \\
		\tau_{z\theta} \\
		Q_{z} %u \tau_{zr} + v \tau_{zz} + w \tau_{z\theta} + q_z
\end{pmatrix}}

\def\Gv{
	\begin{pmatrix}
		0 \\ 
		\mu (u_z + v_r)\\
		\mu (\frac{4}{3} v_z - \frac{2}{3} u_r -  \frac{2}{3} \frac{u}{r})\\
		\mu w_z \\
		u \tau_{zr} + v \tau_{zz} + w \tau_{z\theta} + \zeta T_z
\end{pmatrix}}

\def\Svct{
	\begin{pmatrix}
		0 \\ 
		\tau_{rr} - \tau_{\theta\theta}\\
		\tau_{rz}\\
		\tau_{r\theta}\\
		Q_r %u \tau_{rr} + v \tau_{rz} + w \tau_{r\theta} + q_r
\end{pmatrix}}

\def\Svc{
	\begin{pmatrix}
		0 \\ 
		\frac{2}{r} \mu (u_r - \frac{u}{r})\\
		\frac{1}{r} \mu (u_z + v_r)\\
		\frac{2}{r} \mu (w_r - \frac{w}{r})\\
		\frac{1}{r} (u \tau_{rr} + v \tau_{rz} + w \tau_{r\theta} + \zeta T_r)
\end{pmatrix}}

\begin{equation*}
\textbf{U} = \Uc  \hspace {5mm} \textbf{F} = \Fr 
\end{equation*}

\begin{equation*}
\textbf{G} = \Gz  \hspace {3mm}  \textbf{S}^c = \ScE
\end{equation*}

where total energy $E$ is given by: $ E = \rho e + \frac{1}{2} \rho \textbf{u}^2$. \par 

Assuming a Newtonian fluid, the viscous stress tensor is given by:

\vspace{-3mm}
\begin{equation}
\boldsymbol{\tau} = \mu (\nabla \textbf{u} + \nabla \textbf{u}^T) + \lambda (\nabla \cdot \textbf{u})I
\end{equation}

where $\mu$ is the fluid dynamic viscosity and using stokes hypothesis $\lambda = -\frac{2}{3} \mu$. $\zeta$ is the fluid's thermal conductivity.\par

The flow is assumed to be laminar, and dynamic viscosity is computed via Sutherland's law. Thermal conductivity is computed with constant Prandtl number = 0.71 for the ideal gas case, and $\zeta$ is obtained from the tabulated (precomputed) state data of the EoS \emph{plasma19X} (described in the next section). \par 

Then the viscous flux vectors are derived as follows:

\begin{equation*}
\textbf{F}_v = \Fvt = \Fv
\end{equation*}

\begin{equation*}
\textbf{G}_v = \Gvt = \Gv
\end{equation*}

and the following viscous source term vector results from the cylindrical system conversion:

\begin{equation*}
\textbf{S}^c_v = \Svct = \Svc 
\end{equation*}

Some additional assumptions and resultant simplifications can be made in defining the $\textbf{S}_{MHD}$ vector. \par 

The full expansion of Ohm's law includes a Hall current contribution. This additional current drives the generation of an electric field which has a complex interaction with the gas dynamics - termed \emph{Hall effect}. However, the Hall current can be neglected when the Hall coefficient is sufficiently low. Additionally, several authors have shown that for a conductive body, the Hall effect diminishes the MHD flow control, whereas, in the case of a perfectly insulated body, the overall flow structure is essentially unaffected by the Hall effect \cite{Otsu2010, Fujino2007, Matsuda2008}. Regarding the leading surfaces as an insulated material (via the electrodynamic boundary condition at the body), studies of MHD flow control justifiably neglect Hall effect \cite{Poggie_2002, Khan, Yoshino2009}, therefore applying the \emph{generalised} Ohm's law as given by equation \ref{eq:Ohm_simplified}:

\begin{equation}
\textbf{J}  =	\sigma (\textbf{E} + \textbf{u} \times \textbf{B})
\label{eq:Ohm_simplified}
\end{equation}

%In our methodology, using the simplified Ohm's law for electric current as given in equation \ref{eq:Ohm_simplified}, is especially justified given the relatively low ionisation degree of the LTE air-plasma being examined. \par 

The electric field \textbf{E} is a conservative field where $\textbf{E} = - \nabla \phi$ and $\phi$ is the electric potential. Therefore the governing system is comprised of a system of 5 mixed hyperbolic-parabolic type PDE's as defined within equation \ref{eq:governing_system}, coupled with an additional Poisson-type elliptic PDE maintaining divergence-free current density:

\begin{equation}
\nabla \cdot \textbf{J} = \nabla \cdot [ \sigma (-\nabla \phi + \textbf{u} \times \textbf{B}) ]  = 0
\label{eq:single_elliptic}
\end{equation}

The magnetic field is a static imposed dipole field with dimensional components $\textbf{B} = (B_r, B_\theta, B_z)$, and defined analytically via:

\begin{equation}
\textbf{B} = B_0 \left[ \frac{3\textbf{r}(\textbf{r}\cdot\textbf{m}) - r^2\textbf{m}}{r^5} \right]
\label{eq:B_dipole}
\end{equation}

where \textbf{r} is the radial vector from the dipole centre, and \textbf{m} is an orientation vector aligned parallel to the dipole centreline. Within the axisymmetric governing system where \textbf{r} and \textbf{m} are defined in the r-z plane, the $B_\theta$ magnetic field component is zero. \par  

For electrodynamic boundary conditions which treat the surface as insulated, equation \ref{eq:single_elliptic} can be solved analytically as $\phi = 0$, which means the elliptic current density equation does not require numerical solution and electric current density is given by:

\begin{equation}
\textbf{J} = - \sigma (u_rB_z - u_zB_R) \hat{\theta} = J_\theta
\end{equation}

Therefore the Lorentz forcing terms in the MHD source term vector can be defined directly, rendering the computation as vastly more expedient:

\def\SmhdJB{
	\begin{pmatrix}
		0 \\ 
		(\textbf{J}\times \textbf{B}) \hat{r}\\
		(\textbf{J}\times \textbf{B}) \hat{\theta}\\
		(\textbf{J}\times \textbf{B}) \hat{z}\\
		\textbf{u} \cdot (\textbf{J} \times \textbf{B})  + \sigma^{-1} J^2
\end{pmatrix}}

\def\Smhd{
	\begin{pmatrix}
		0 \\ 
		- \sigma (u_rB_z^2 - u_z B_rB_z)\\
		0\\
		\sigma (u_rB_zB_r - u_z B_r^2) \\
		u_r J_\theta Bz - u_z J_\theta B_r + \frac{1}{\sigma} J^2
\end{pmatrix}}

\begin{equation*}
\textbf{S}_{MHD} = \SmhdJB = \Smhd
\end{equation*}

\subsection{Equation of state for air-plasma}

The system equations are closed with an advanced 19 species equilibrium air-plasma equation of state (EoS), applied for the first time to applications in hypersonically generated air plasmas. The new 19 species approach to the thermochemistry modelling is an extension to the standard 11-species model \cite{CEA}, and includes a direct and accurate computation the electrical conductivity transport property, which has previously posed challenges in MHD studies \cite{Bisek}. \par 

The advanced EoS considers 19 species reaching chemical and thermal equilibrium, and therefore can be applied to aerospace plasma problems determined to be in local thermodynamic equilibrium (LTE). The EoS is an extension of the precomputed \emph{plasma19} database originally developed by Tr\"{a}uble in 2018 \cite{Trauble2018} (which utilised key theory from D'Angola \cite{DAngola1} and Colonna \cite{Colonna2007})), and as published in 2021 for the application of simulating air-plasma generated during lightning strike \cite{Trauble2021}. Extensions were made to this database in this work, based on the original theory but in order to generate properties over a wider density-pressure range necessary for the simulation of air-plasma generated by hypersonic flows at the low pressure and density freestream conditions typical of this regime. Additionally, an extension to the computation of electron number density was implemented, following the theory of D'Angola \cite{DAngola2}. The new extended EoS will therefore be referred to as \emph{plasma19X}. 

The 19 species considered in the air-plasma are: \par 

\begin{center}
	$N_2$, $N_2^+$, $N$, $N^+$, $N^{++}$, $N^{+++}$, $N^{++++}$, \\
	$O_2$ , $O_2^+$, $O_2^-$, $O$, $O^-$, $O^+$, $O^{++}$, $O^{+++}$, $O^{++++}$, \\
	$NO$, $NO^+$, $e^-$
\end{center}

The development of the 19 species EoS considers a large number of basis chemical reactions, where every such reaction is governed by a law of mass action corresponding to dynamic equilibrium, and total mass is conserved across all species. For a specified intensive property pair (e.g. pressure, temperature) a large, determined system of non-linear equations governs the precise species composition. Numerical solution for species composition, and resultant thermodynamic properties from detailed curve fits which follows the work by D'Angola \cite{DAngola1}. 
%An accurate hierarchical algorithm for computation of chemical composition at LTE of a given reactive mixture has been developed by D'Angola \cite{DAngola1}. It achieves the computationally difficult task of solving the large set of system equations containing equilibrium constants of widely varying orders of magnitude. Once the chemical composition is determined, the thermodynamic properties and potentials (Gibbs free energy, total free energy, total internal energy and enthalpy) can be ascertained from the relevant partition functions. Detailed numerical fits have been developed for the computed properties. 
%For transport properties of the system (such as thermal and electrical conductivity), the equations are based on Chapman-Enskog theory, a detailed derivation of which can be found in Capitelli et al. \cite{Capitelli2012}. 
For the remainder of the specifics of the developed theory for plasma19, please refer to the detailed description in the documented work\cite{Trauble2018}.

Following the theory of plasma19's development, thermodynamic properties are physically extended over the low pressure and temperature range required for freestream conditions of hypersonic flight cases. A method to very accurately compute electron number density down to very low air plasma temperatures was implemented for the extended tabulated plasma19X EoS following the extended theory proposed by D'Angola \cite{DAngola2}. This is critical to the plasma properties of the temperature band of interest in hypersonically formed plasmas (approximately 2,000-10,000 K). The extended plasma19X EoS can compute thermodynamic properties, transport properties and species composition for a density range of 10$^{-5}$  kg/m$^3$ - 10 kg/m$^3$ and a pressure range of 7 Pa to 18.23 MPa, which corresponds to a temperature band of approximately 10 K - 60,000 K. The state properties computed by plasma19X have been extensively validated against empirical measurements \cite{Trauble2021}.

\subsection{Electrical conductivity model}

%This work demonstrates improvements in the prediction of electrical conductivity in the temperature band of interest for hypersonically formed equilibrium air plasma. 

Accurate modelling of any magnetogasdynamic effect is underpinned by the accuracy of the computed electrical conductivity. This is because conductivity drives the Lorentz forcing dynamics of the flow:

\begin{equation}
\textbf{J} \times \textbf{B} = \sigma \cdot [\textbf{u} \times \textbf{B}] \times \textbf{B}
\end{equation}

As Bisek \cite{Bisek} highlights, accurate conductivity modelling typically involves semi-empirical approximations and remains problematic in MHD studies. 

Ideal gas models (widely used in MHD studies \cite{Otsu2010} due to the computational efficiency) cannot inherently compute electrical conductivity, and so, adopt a highly simplified supplementary model. The supplementary conductivity model is typically an analytic power function of temperature \cite{Poggie_2001, PorterCambel, Dias_MHD}:

\begin{equation}
\sigma = \sigma_0 \left(\frac{T}{T_0}\right)^n
\label{eq:finite_conductivity}
\end{equation}

where constants $\sigma_0$ and $T_0$ are manually fitted to the problem, and exponent $n$ takes a value in the range: 0-4 depending on the test case. 

\begin{figure}[h!]
	\centering
	\includegraphics[width=8cm]{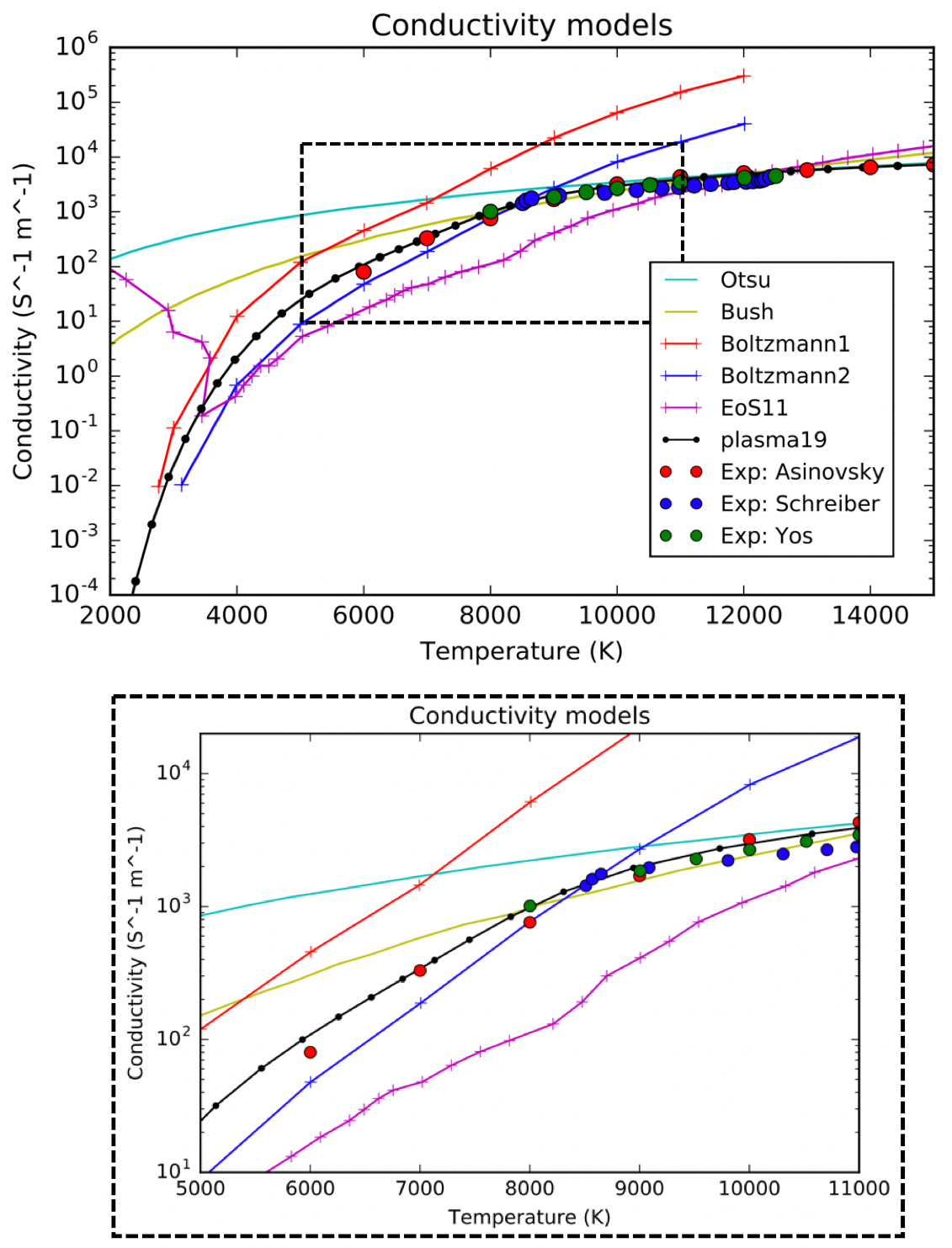}
	\caption{Electrical conductivity, at pressure = 1 atm, computed by plasma19X, compared with other models currently used in the literature and with experiment. Three sets of experimental measurements have been obtained from the works of Asinovsky et al. (1973) \cite{Asinovsky1971}, Schreiber et al. (1971) \cite{Schreiber1973} and Yos (1963) \cite{Yos1963} where a low-temperature equilibrium plasma was generated via a stabilised DC arc applied to a column of air maintained at atmospheric pressure.  Otsu \cite{Otsu2010} and Bush \cite{Bush1958} use the simple analytic power law approximation defined in equation \ref{eq:finite_conductivity} with exponents 2 and 4 respectively. Solution to the Boltzmann equation is as per the solver developed by Weng and Kushner \cite{WengKushner1990} for two different levels of resolution. EoS11 is an 11-species equilibrium model as implemented by Villa et al. \cite{Villa1}. }
	\label{fig:conductivity_models}
\end{figure}

Real gas models (both equilibrium and non-equilibrium) also adopt supplementary conductivity models which are functions of electron number density ($N_e$), and therefore critically depend on $N_e$ as computed by species dissociation and ionisation of the full thermochemistry model. A standard form for weakly ionised equilibrium plasmas is given by:

\begin{equation}
\sigma = \frac{N_e e^2}{m_e \nu_{coll}} %, \hspace{3mm} \nu_{coll} = \frac{5.891 \times 10^{17} \cdot p }{T^{1/2}}
\end{equation}

where $n_e$ is the electron number density, $e$ is the electron charge, $m_e$ is the electron mass, and $\nu_{coll}$ is the effective electron-neutral and electron-ion collision frequency. Several different formulations for $\nu_{coll}$ (where the gas is assumed to follow the Maxwell Boltzmann distribution) are commonly considered \cite{Raizer1991}. These are typically in the form of analytic functions of pressure and or temperature, where constants vary depending on the empirically identified regime \cite{Bisek}.  

The electrical conductivity given by plasma19X, however, was computed by the detailed theory of perturbative Chapman-Enskog method, which derives a full set of collision integrals from the species compositions, and a third order approximated diffusion coefficient ($D_{ee}(3)$). which then computes electrical conductivity as:

\begin{equation}
\sigma = e^2\frac{N_e m_e n}{\rho k_B T} D_{ee}(3)
\label{eq:sigma19}
\end{equation}

where $n$ is the total number density and $k_B$ is the Boltzmann constant. The detailed derivation can be found in Capitelli et al. \cite{Capitelli2000}. 

FIG. \ref{fig:conductivity_models} demonstrates the accuracy of the computed $\sigma$ by plasma19X as compared with experiment and other commonly employed supplementary models. As can be seen in FIG. \ref{fig:conductivity_models}, the non-physical species molar fractions over the low temperature range, and resultant anomalous electrical conductivity, render EoS11 as problematic for LTE MHD studies. All models must be both smooth and accurate to produce stable and realistic computational results for MHD studies. While different models perform well in different temperature bands, this is often achieved by manually fixing parameters to the computed temperature and profile of the flow. Plasma19X is found to be smooth and accurate (without any manual input), especially over the temperature band of interest for hypersonically formed plasmas in the equilibrium regime.

The accuracy of plasma19X can be attributed to the detailed method of derivation of $\sigma$ in equation \ref{eq:sigma19} including the additional species dissociations and ionisations captured at relatively low air plasma temperatures in the 19 vs 11 species approach. With relevance to the application of this work, the double and triple ionised species captured by plasma19X are only present at higher temperatures (>20,000 K), and largely do not effect the state in the temperature band of interest (approximately 2,000 K - 10,000 K). The negatively charged $O^-$, $O_2^-$ and $N_2^-$ species however, are not captured by an 11-species model but are computed to be present in small concentrations in the low air plasma temperature range. The implementation of electron molar fraction is accurate down to temperatures as low as 50 K as compared with verified result of D'Angola \cite{DAngola2}. 

Plasma19X also has the advantage that $\sigma$ is obtained directly via the pre-computed tabulated EoS, rendering the model as computationally efficient. 

\section{Numerical model}

\subsection{Mesh generation and geometry implementation}

\begin{figure}[h!]
	\centering
	\includegraphics[width=7cm]{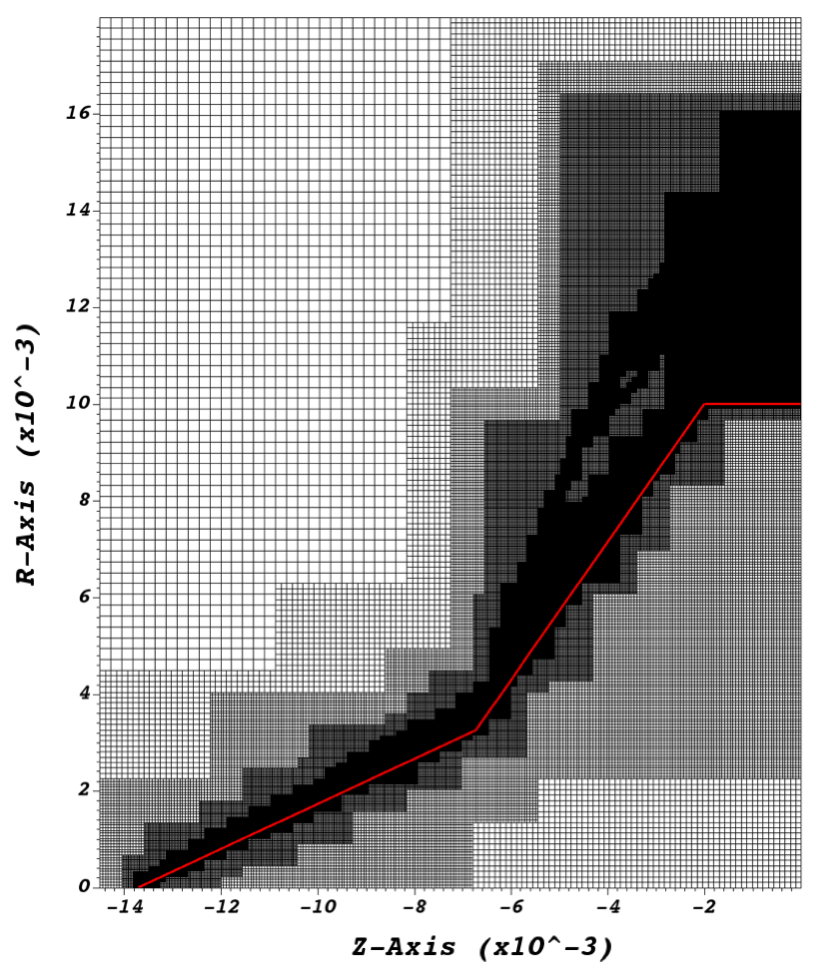}
	\caption{Cartesian mesh configuration with four levels of hierarchical AMR shown for an example mesh underlying a double cone steady state solution. Embedded boundary of the rigid body geometry is shown in red.} 
	\label{fig:DoubleConeMesh}
\end{figure}

The numerical model of this work is developed within the AMReX framework \cite{amrex}. AMReX provides a platform for hierarchical adaptive mesh refinement (AMR) and massively parallel computation of domain sub-blocks. All numerical methods were implemented directly \emph{within} this framework in developing the model of this work. The hierarchical AMR (depicted in FIG. \ref{fig:DoubleConeMesh}) and subdivision of the regular Cartesian mesh, which covers the full domain, results in efficient and highly scalable computation. 

Due to the full domain discretisation of the AMReX framework, geometry is implemented via an embedded boundary method. A sharp interface rigid body Ghost Fluid Method (GFM) was implemented, such as that employed by Bennett et al. \cite{Bennett2018}, based on the approach of Sambasivan and UdayKarmar  \cite{Sambasivan}. 

By implementing embedded boundary geometry via a rigid body GFM, the mesh is not subject to distortions and skewing which sometimes occur in the case of body fitted meshes, especially as geometries become more complex. The Riemann-based GFM is robust for very strong shocks which interact with the embedded surface. The adaptive refinement approach ensures high resolution is achieved automatically and locally in the vicinity of high density gradients and at the body-fluid boundary. This combination of methods therefore permits the simulation of non-simple geometries, in the presence of strong interacting shocks, with multiple levels of hierarchical AMR applied to achieve high effective resolution and solved efficiently via MPI-based computation with high scalability across many nodes.\par

\subsection{Numerical solver methodology}

The combination of algorithms, implemented within the AMReX framework, is a recently developed methodology, outlined in a previous publication \cite{Muir2021}, designed for the efficient simulation of hypersonic MHD systems. 

The inviscid system fluxes are evaluated numerically via a complete wave approximate Riemann solver (HLLC) with high order extension (MUSCL) which is second order accurate in space and time. Viscous derivative terms are evaluated using the central differencing method at cell interfaces with divergence then computed through the finite volume construction. The transient solver can be used for time accurate simulation of transient test cases, and also reaches a stable and convergent steady state, for such test cases.

The system is solved via an operator split strategy whereby the hyperbolic and diffusive fluxes are computed successively with the result of each time-wise integration input as the initial condition for the subsequent operation. Fluxes are also dimensionally split in this manner. Cylindrical source terms are integrated via Strang splitting to maintain second order accuracy.

\begin{figure*}[ht]
	\centering
	\includegraphics[width=17cm]{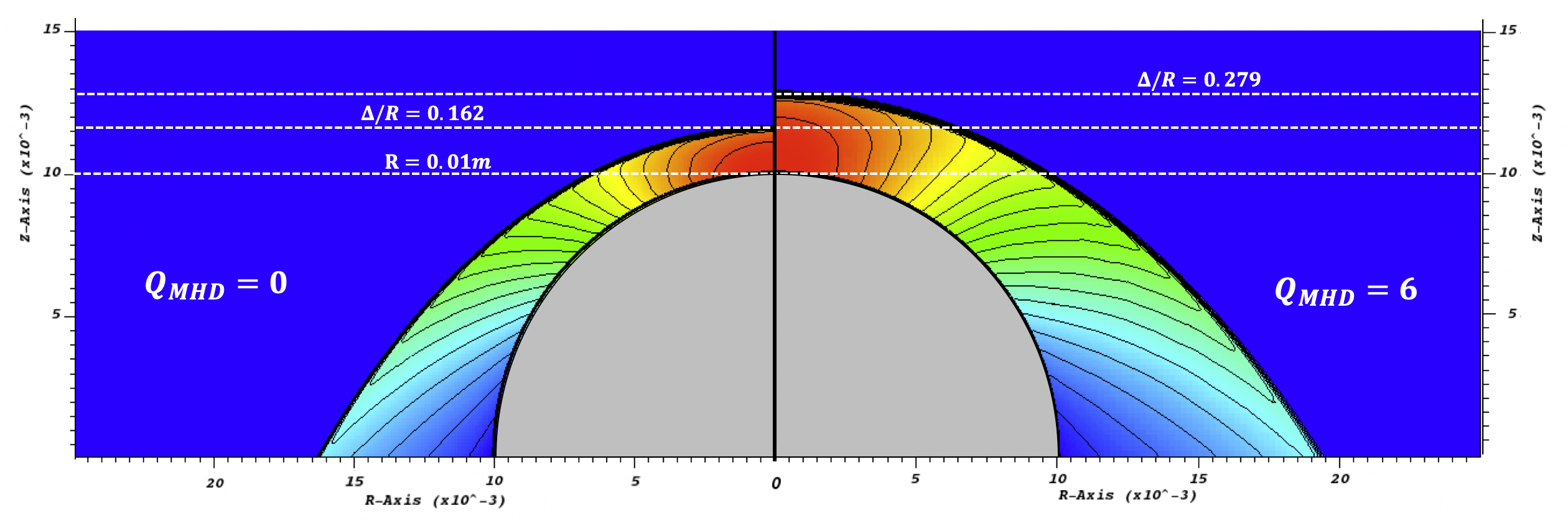}
	\caption{Pressure profiles with contours as computed in this work for $Q_{MHD} = 0$ and $Q_{MHD} = 6$.} 
	\label{fig:pressure_validation}
\end{figure*}

The Lorentz forcing ODEs of the MHD model can be expressed in the condensed form:

\def\UODE{
	\begin{pmatrix}
		\rho \textbf{u} \\
		E \\ 
	\end{pmatrix}
}

\def\SODE{
	\begin{pmatrix}
		\textbf{J} \times \textbf{B} \\
		\textbf{u} \cdot (\textbf{J} \times \textbf{B}) + \eta J^2 \\ 
	\end{pmatrix}
}

\begin{equation}
\frac{\partial}{\partial t} \UODE = \SODE
\end{equation}

where $\eta = \sigma^{-1}$ is the plasma resistivity. An explicit Euler integration approach is adopted, where an additional momentum half-step is utilised:

\begin{align}
(\rho \textbf{u})^{n+\frac{1}{2}} &= (\rho \textbf{u})^{n} + \frac{\Delta t}{2} \cdot (\textbf{J}^n \times \textbf{B}) \label{eq:half_mom_integration} \\
(\rho \textbf{u})^{n+1} &= (\rho \textbf{u})^{n+\frac{1}{2}} + \frac{\Delta t}{2} \cdot (\textbf{J}^n \times \textbf{B}) \label{eq:mom_integration} \\
E^{n+1} &= E^n + \Delta t \cdot \left(\textbf{u}^{n+\frac{1}{2}} \cdot  (\textbf{J}^n \times \textbf{B}) + \eta^n (J^{n})^{2} \right) \label{eq:en_integration}
\end{align}

The momentum half-step, and resultant $\textbf{u}^{n+\frac{1}{2}}$ are carefully constructed to ensure the $\textbf{J} \times \textbf{B}$ specifically effect momentum whilst conserving internal energy. The joule heating term $\eta^n (J^{n})^{2}$ only acts upon system internal (thermal) energy. See full derivation in Villa et al. \cite{Villa1}. 

Some careful treatment of system stability is required for dimensionally and operator split systems when computed within an AMR framework. 

The stable time-step is restricted as the minimum of the hyperbolic and diffusive stable time steps, given by: 

\begin{align}
\Delta t &= \min[\Delta t_{hyp}, \Delta t_{diff}] \\
&= \min[\frac{\text{CFL} \cdot \Delta x_{d,i}}{S_{max,i}} , \frac{(\Delta x_{d,i})^2}{2 \cdot \max[\frac{\mu_i}{\rho_i}, \frac{\zeta_i}{(\rho c_p)_i}]}] 
\end{align}

where $\Delta x_{d,i}$ is each dimensional spatial step size. $\Delta t_{diff}$ is based on the formulation recommended by Gokhale \cite{GokhalePhD2018}. Under the operator split solution strategy, the explicit time-marching evolution is stable in 2D (including all source terms) for CFL=0.9. However, stable time-step is computed on the $0^{th}$ (coarsest) level of the Cartesian mesh, with higher levels sub-cycled with matched integer fractions of the base-level time-step. Therefore, in consideration of multiple AMR levels and the combined hyperbolic-parabolic nature of the Navier-Stokes system equations, the stable time-step as computed at the $0^{th}$ mesh level is:

\[
\Delta t = 
\begin{cases}
\Delta t_{diff}, & \text{if }  \left(\Delta t_{diff} \cdot (\frac{1}{2})^{2 \cdot \text{N\_MAX}}\right)   <    \left( \Delta t_{hyp} \cdot (\frac{1}{2})^{\text{N\_MAX}} \right) \\
\Delta t_{hyp},              & \text{otherwise} 
\end{cases}
\]
\\

where N\_MAX is the highest AMR level and a refinement factor of 2 is assumed. 

\section{Validation tests}

A large number of tests have been performed to extensively validate the new numerical model. A set which collectively validates the key physics of this study is presented here. 

\subsection{Low $Re_m$ MHD system}

The accepted benchmark validation test for a hypersonic MHD system under low magnetic Reynolds number assumption, is the theoretical test originally proposed by Poggie \& Gaitonde \cite{Poggie_2001}. It has been replicated subsequently by Damevin \& Hoffmann \cite{Damevin2002} and most recently by Fujino et. al \cite{Fujino2008}. For the purpose of validating the underlying resistive MHD model, the model of this work is reduced to its inviscid Euler equivalent, closed with an ideal gas equation of state for consistency with this validation test cases in the literature. \par 

The free-stream conditions for the flow over a 10 mm radius sphere are: \par 

\begin{center}
	\underline{Free-stream conditions:} \\
	M = 5, Re = 80,000, $T_\infty$ = 100 K, \\
	$V_\infty$ = 1002.25 m/s, $\rho_\infty$ = 0.0798 $kg/m^3$, $p_\infty$ = 2290.85 Pa \\ 
	Gas = Ideal Air, $\gamma = 1.4 $ 
\end{center}

For the simplified ideal gas model, the electrical conductivity is given by the analytic power law of equation \ref{eq:finite_conductivity}, where $\sigma_0 = 300 \text{ } (\Omega m)^{-1} $ and n=0 which therefore sets a constant post-shock conductivity through the shock layer (initiated via the condition: $T > T_\infty$). $B_0$ is the maximum absolute magnetic field strength outside of the vehicle, and the characteristic length $R_0$ is the radius of the spherical body. \par 

The MHD enhancement effect is shown to increase shock stand-off distance with increased $Q_{MHD}$ as defined in Equation \ref{eq:QMHD}. For this test $Q_{MHD}$ is varied from 0-6 by corresponding maximum \textbf{B}-field strength. 

\begin{figure}[h!]
	\centering
	\includegraphics[width=8.5cm]{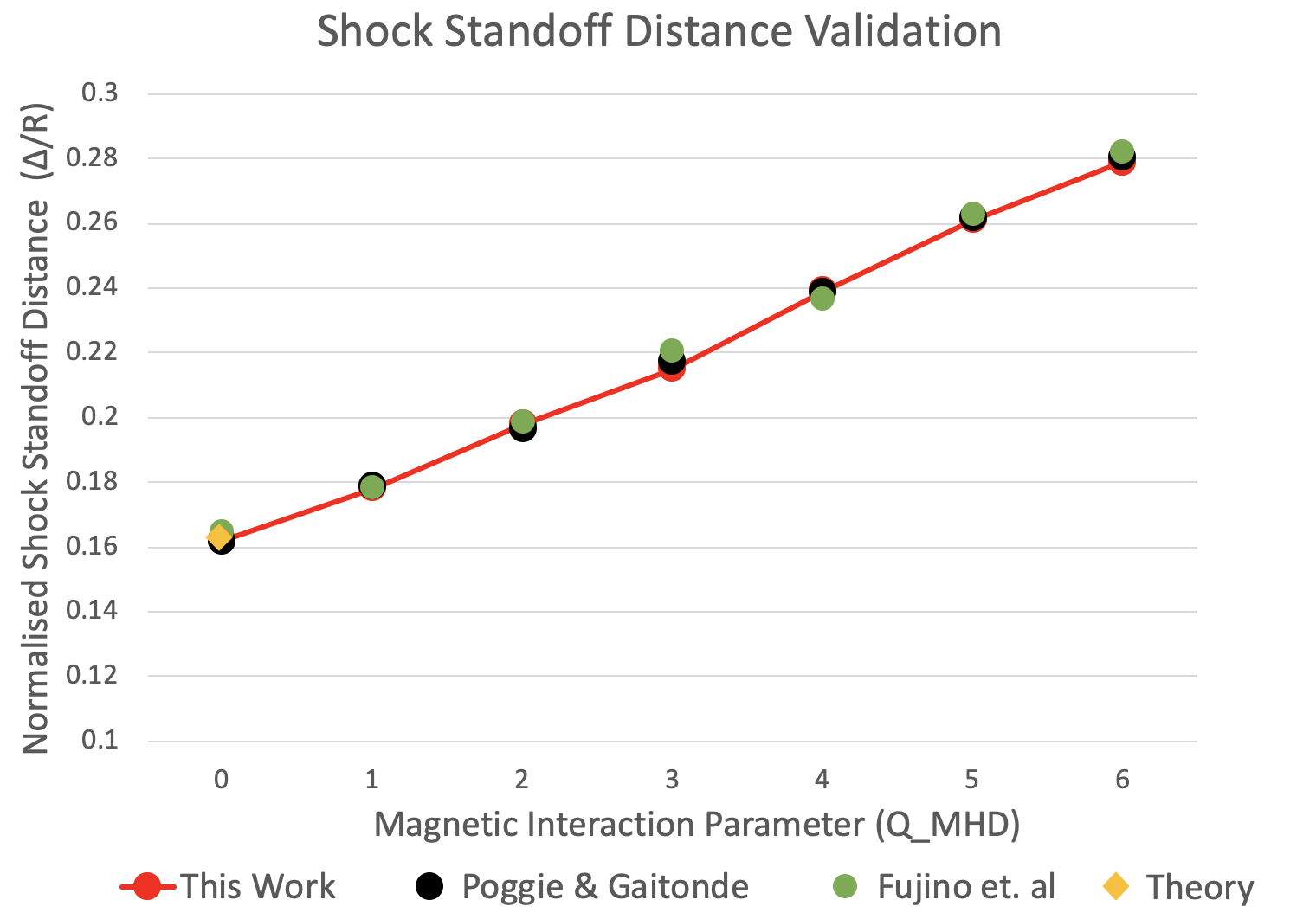}
	\caption{Quantitative comparison of shock stand-off distance for this work vs previous studies for $Q_{MHD} = 0-6$.} 
	\label{fig:standoff_validation}
\end{figure}

The full pressure profile for the $Q_{MHD} = 0$ and $Q_{MHD} = 6$ cases are shown in FIG. \ref{fig:pressure_validation} and are in very good agreement with the literature. Examining the result quantitatively, FIG. \ref{fig:standoff_validation} shows that the shock stand-off distances computed in this work under the different magnetic field strengths exhibit excellent agreement with literature results and theory. \par 

\FloatBarrier 

\subsection{Hypersonic double cone experiments}

The hypersonic double cone tests are a relevant test for the validation of Navier-Stokes solvers where non-simple geometries produce SWBL interactions. Experimental data has been obtained from a large number of tests in the CUBRC LENS expansion tube facilities \cite{Holden2002} producing a high enthalpy, hypersonic, laminar flow with shock wave boundary layer interactions. \par 

\begin{figure}[h!]
	\centering
	\includegraphics[height=6.0cm]{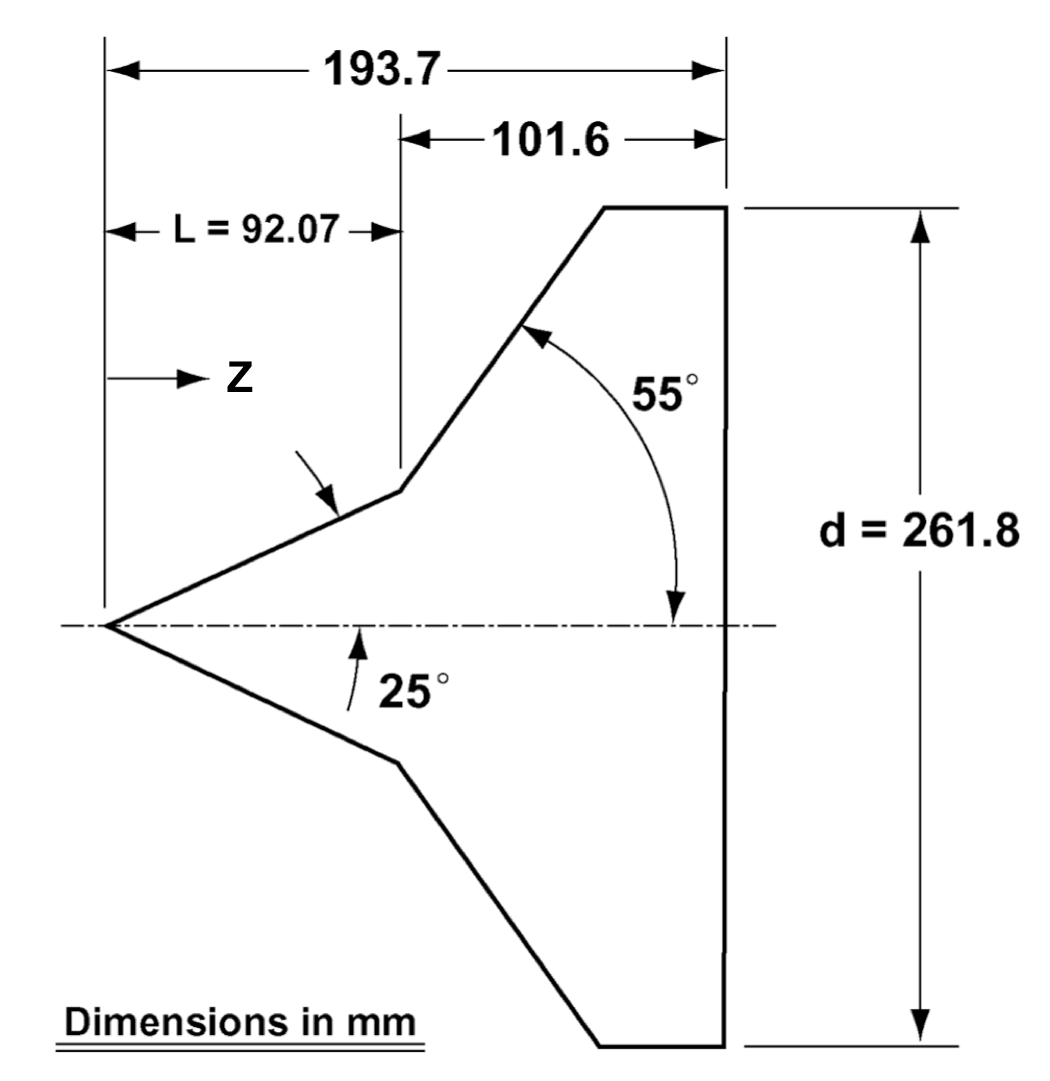}
	\caption{Geometry specifications of the CUBRC 25/55$^o$ double cone model} 
	\label{fig:CUBRC_dimensions}
\end{figure}

%\begin{figure}[h!]
%	\centering
%	\includegraphics[height=7.3cm]{./vis/Schematic_TypeA.png}
%	\caption{Schematic of the known shock structure formed \cite{CUBRC2014} for a hypersonic flow over the CUBRC double cone geometry. All flow features are labelled, including sonic lines to depict the supersonic and subsonic regions of the flow. This flow structure is classified as TYPE A in this paper.} 
%	\label{fig:Schematic_TypeA}
%\end{figure}

\begin{figure}[h!]
	\centering
	\includegraphics[height=8.0cm]{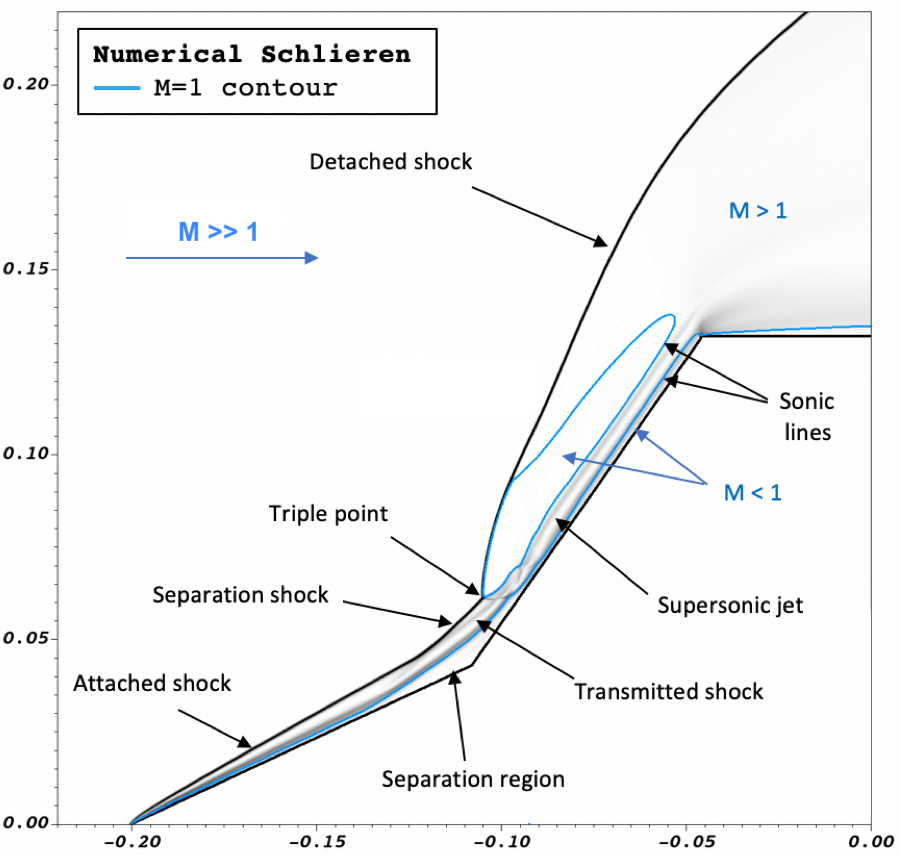}
	\caption{Numerical Schlieren result of this work for CUBRC test conditions, showing complete feature formation as compared with the known flow structure \cite{CUBRC2014}.} 
	\label{fig:DoubleConeSchlieren}
\end{figure}

FIG.  \ref{fig:CUBRC_dimensions} shows the model dimensions for the CUBRC experiments, and FIG. \ref{fig:DoubleConeSchlieren} shows a numerical Schlieren result from the model of this work for a typical CUBRC test condition. The numerical Schlieren of FIG. \ref{fig:DoubleConeSchlieren} demonstrates the ability of the Navier-Stokes rigid body GFM model to capture the complete feature formation known to be produced at steady state for this flow test (all features are identified to match prevailing work \cite{CUBRC2014, Hao2017}). 

%A flow structure classification system is designated in this paper to aid the analysis of section \ref{sec:analysis}. The CLASS A schematic of FIG. \ref{fig:Schematic_Types} represents the identified shock structure and all labelled features which are known to form for the test runs of the CUBRC double cone experiments \cite{CUBRC2014}. Comparison of the numerical Schlieren of FIG. \ref{fig:DoubleConeSchlieren} with the CLASS A schematic demonstrates the ability of the Navier-Stokes rigid body GFM model to capture the complete feature formation known to be produced at steady state. 

To validate the system quantitatively a pressure trace is computed along the surface of the vehicle and compared with the results of the experiments and other simulations. Test run 7 is typically chosen due to the availability of smooth experimental data, and results of multiple numerical studies for comparison. Since the test gas is non-reactive nitrogen ($N_2$), the ideal gas EoS was used to close the system. 

\begin{center}
	\underline{CUBRC run 7 conditions:} \\
	M = 15.6, $V_\infty$ = 2073 m/s, $T_\infty$ = 42.6 K, $T_{w}$ = 300 K,\\ 
	$\rho_\infty$ = 1.57 $\times 10^{-4}$ $kg/m^3$, $p_\infty$ = 2.23 Pa, \\
	Gas = $N_2$, $\gamma = 1.4$ 
\end{center}

\begin{figure}[h!]
	\centering
	\includegraphics[width=8.5cm]{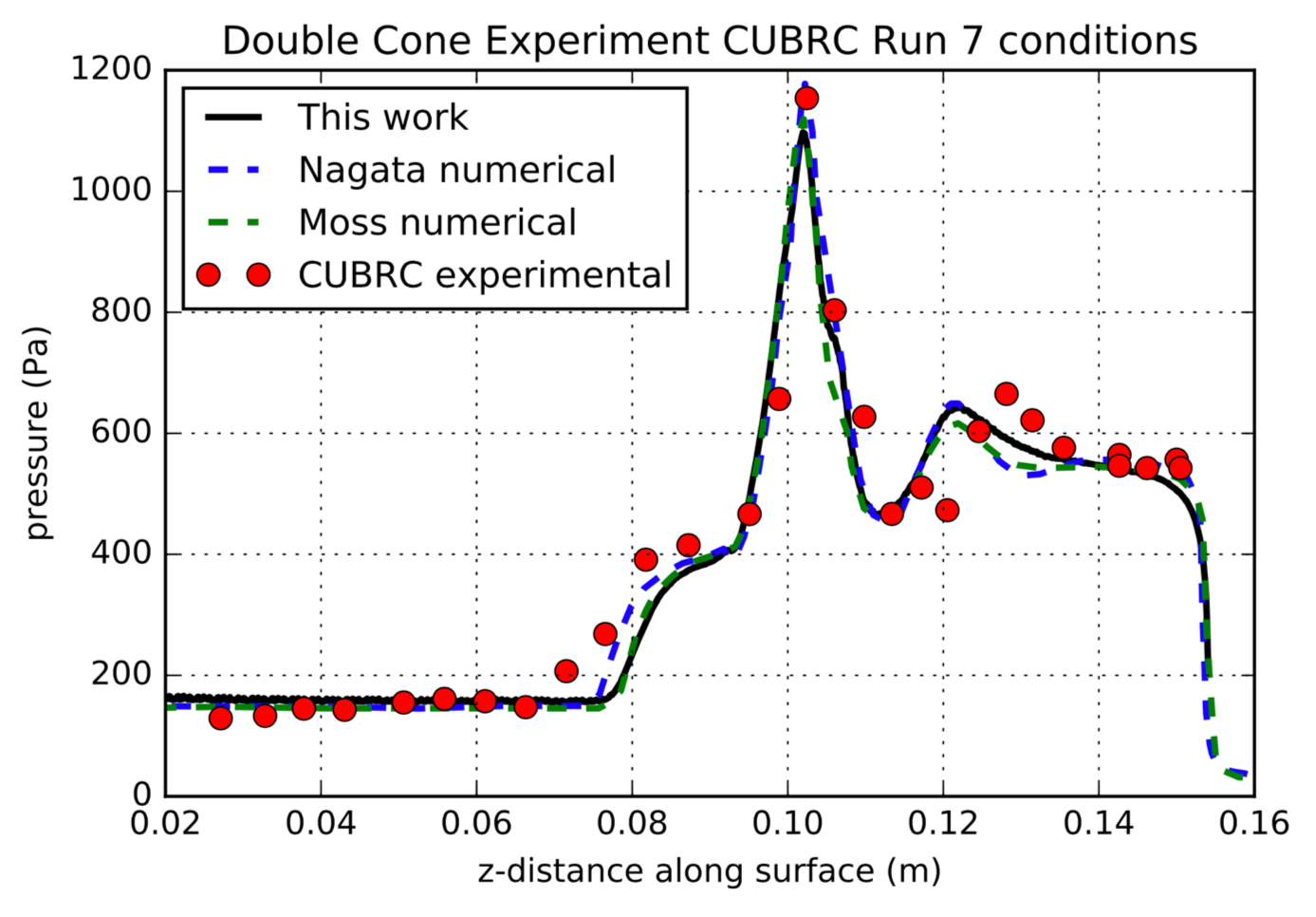}
	\caption{Pressure trace along surface of double cone: left: CUBRC run 7 result of this work compared with experimental result and simulations of Nagata et al. \cite{Nagata2011numerical} and Moss et al. \cite{Moss2005}} 
	\label{fig:CUBRC7}
\end{figure}

Very good agreement between the simulation, experiment, and the results of other numerical models is demonstrated in Figure \ref{fig:CUBRC7}.

\FloatBarrier

\subsection{MHD hypersonic double cone experiment}
\label{sec:MHD_exp_val}

Wasai et al. \cite{Wasai2010} conducted a seminal experiment on magnetohydrodynamic control of shock interactions over a $25^o/55^o$ double cone geometry in hypersonic flow where results were obtained to measure the MHD flow control effect on shock enhancement. Simulations of the experimental conditions have been conducted by Nagata et al. \cite{Nagata2011numerical}$^,$\cite{Nagata2013Bfield}. \par 

\begin{figure}[h!]
	\centering
	\includegraphics[width=9.0cm]{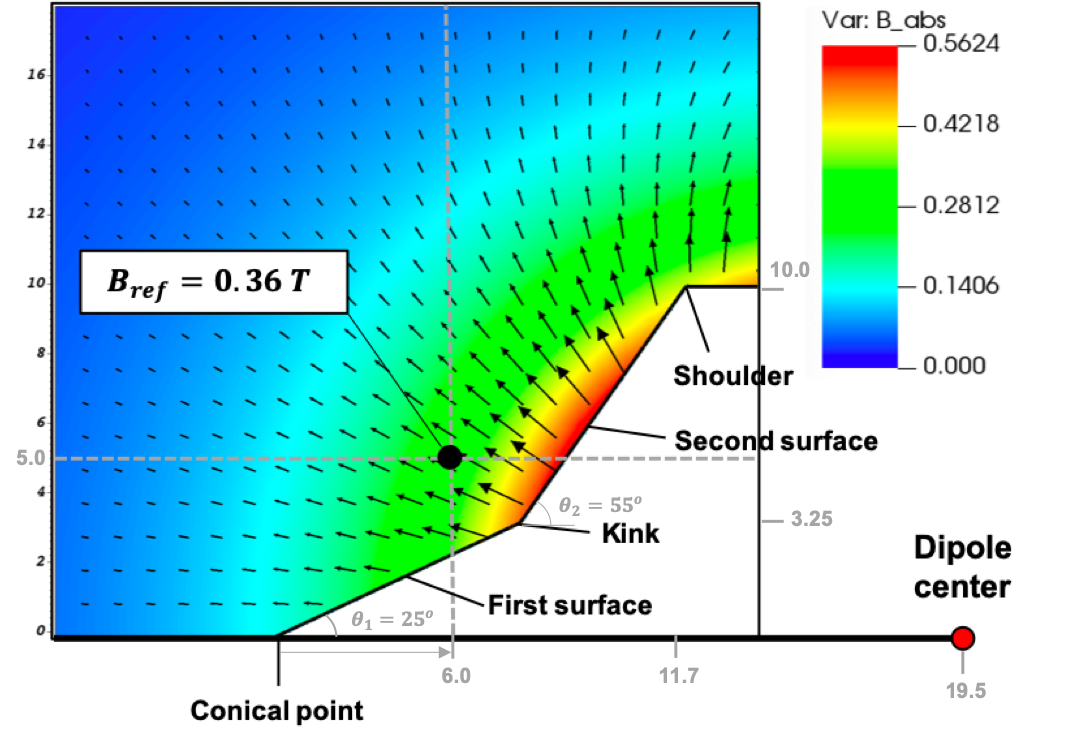}
	\caption{Geometry specifications and computed magnetic field matched with the experimental conditions of Wasai et al. \cite{Wasai2010} with dimensions in mm. Key terms labelled. The case of B=0.36 T is the measured magnetic field strength at the reference point.} 
	\label{fig:Bconegeometry}
\end{figure}

The problem configuration is congruent to the CUBRC double cone experiments with model dimensions shown in FIG. \ref{fig:Bconegeometry}. A dipole magnetic field is initialised from within the model, with the dipole centre depicted in the figure, and a magnetic field strength of B=0.36 T measured at the labelled reference point. 

\begin{figure}[ht]
	\centering
	\includegraphics[width=8.0cm]{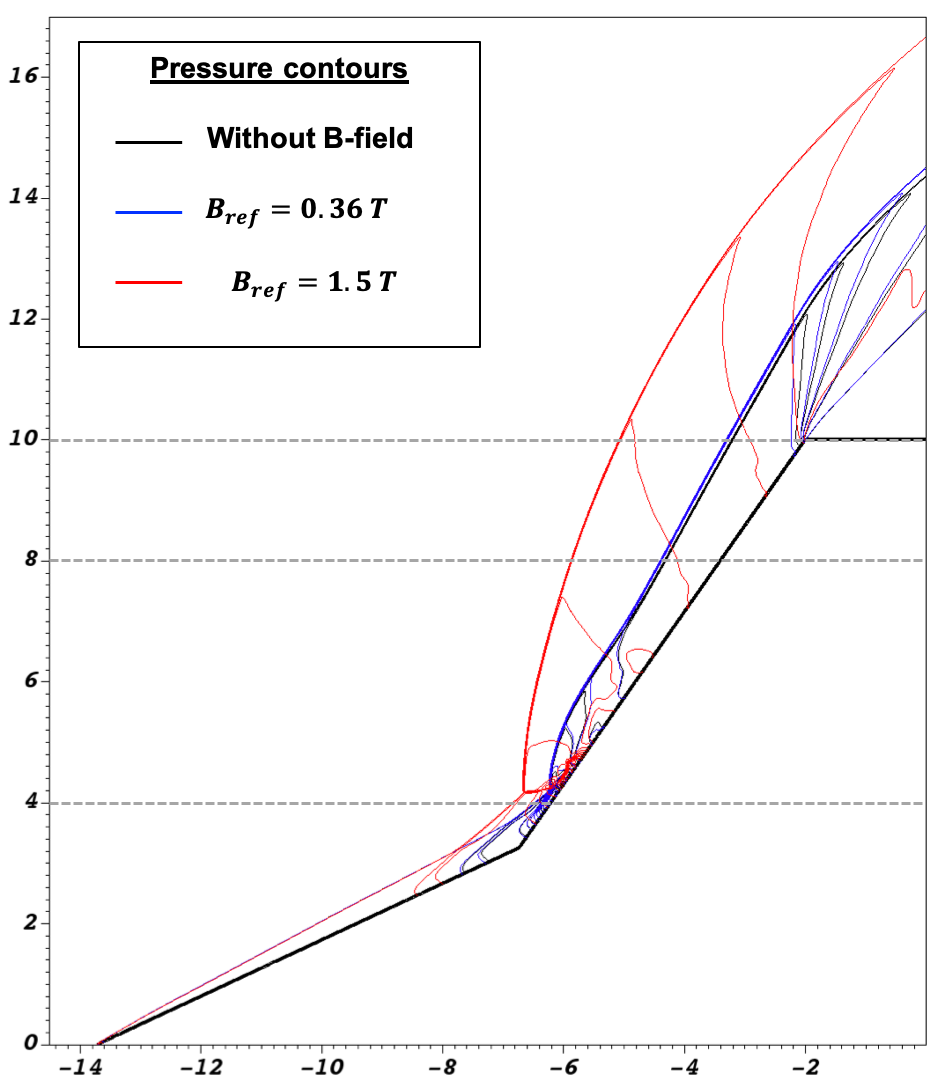}
	\caption{Contours of pressure for the cases: without magnetic field, magnetic field initialised to experimental value $B_{ref} =0.36$ T, and a stronger magnetic field of $B_{ref} = 1.5$ T intensity. Radial lines (along which property traces are taken) of 4 mm, 8 mm, 10 mm are marked.} 
	\label{fig:MHDpressurecontours}
\end{figure}

The experimental inflow test gas was pre-heated, and Wasai \cite{Wasai2010} and Nagata \cite{Nagata2013Bfield} determined it to have reached a state of thermochemical equilibrium, calculating the initial (single) temperature and species composition as such. Of the 18 test runs conducted, flow velocities were measured between 11.4-12.5 km/s \cite{Wasai2010}, with a nominal value of 12.1 km/s used for simulations. 

%Experimental results for the test run of flow velocity of 11.6 km/s are most clearly presented by Wasai and therefore used as the test flow condition for this work. 

\begin{center}
	\underline{Test run conditions:} \\
	M = 5.6, $V_\infty$ = 12.1 km/s, $T_\infty$ = 6110 K, $T_{w}$ = 300.0 K,\\ 
	$\rho_\infty$ = 2.52 $\times 10^{-3}$ $kg/m^3$, $p_\infty$ = 7.22 kPa, Gas = Air
\end{center}

The flow is characterised with the following parameters:

\begin{center}
	Reynolds number = 1.77 $\times 10^3$ \\
	Magnetic Reynolds number = 0.2 \\
	Hall parameter = 0.61
\end{center}

These parameters confirm the assumptions of the governing mathematical model: low magnetic Reynolds number, and negligible Hall effect (since additionally the model surface can be electrically insulative \cite{Wasai2010}). The Nagata et al. model makes the same assumptions \cite{Nagata2011numerical}. \par 

To show the effect of increasing the magnetic field strength (as per the Nagata simulations), a simulation is run with a stronger \textbf{B}-field: fixed at the same dipole centre location but scaled to a larger magnitude corresponding to 1.5 T at the reference point. \par  

The MHD effects are depicted by the overlaid pressure contours for different magnetic field strength in FIG. \ref{fig:MHDpressurecontours}. The small but clearly observable magnetic augmentation of the shock position in the B = 0.36 T case demonstrates the importance of high resolution numerical methods to sharply capture the shock wave position.

\begin{figure*}[ht]
	\centering
	\includegraphics[width=17.0cm]{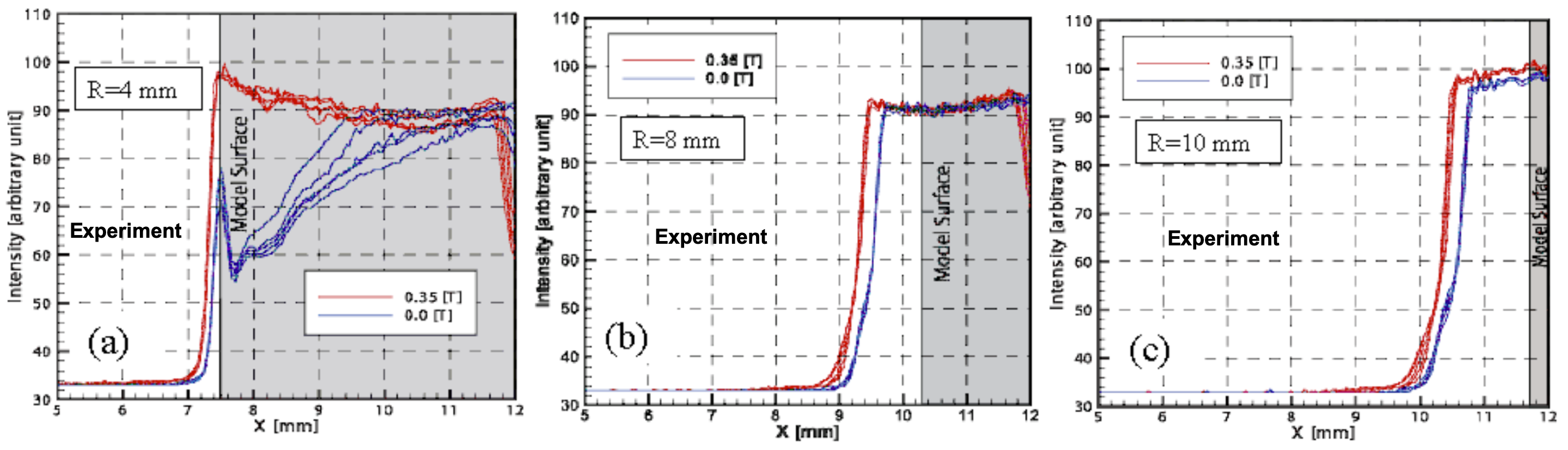}
	\includegraphics[width=17.0cm]{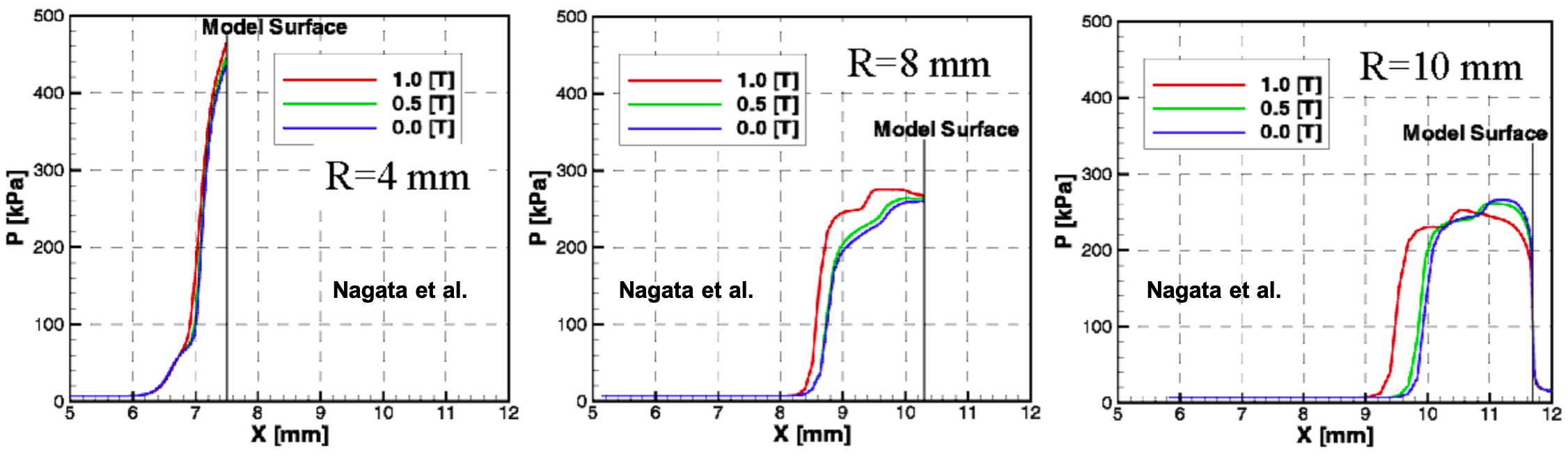}
	\includegraphics[width=17.0cm]{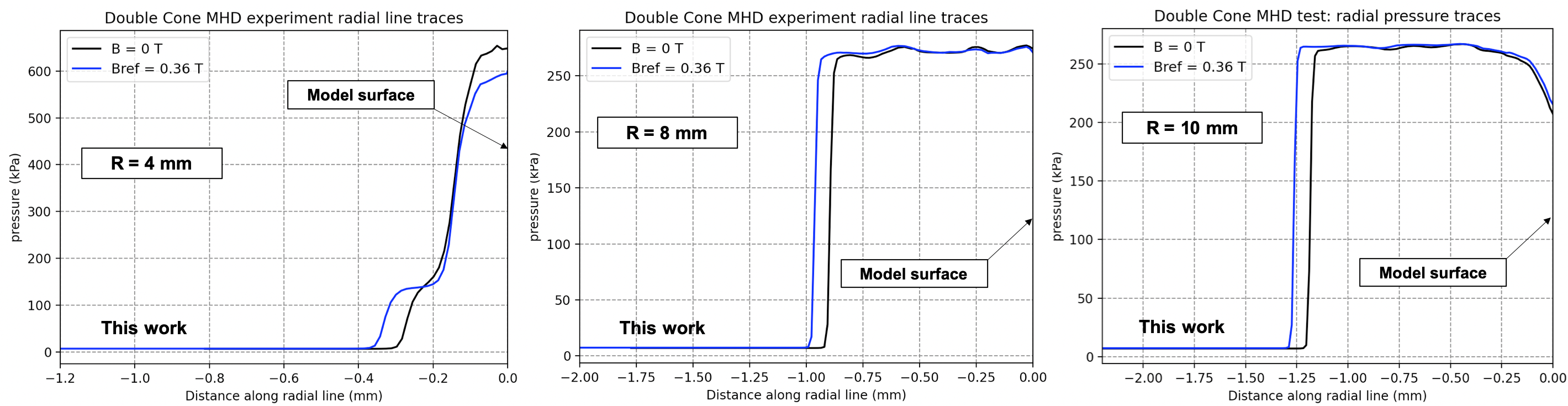}
	\caption{Top) experimental result brightness intensity traces through radial lines 4 mm, 8 mm, and 10 mm, for a number of test runs by Wasai et al.\cite{Wasai2010}. Middle) pressure traces taken at the same radial lines from the numerical result of Nagata et al. \cite{Nagata2011numerical}. Bottom) pressure traces taken at the same radial lines from the numerical result of this work. Experimental and Nagata traces are presented as total distance along the central z-axis (labelled X-axis in the experimental results), and simulation traces are shown more clearly as distance along the z-axis measured directly from the surface intersection point as z=0 for each case.} 
	\label{fig:radial_traces} 
\end{figure*}

\begin{table}
	\centering
	\caption{Comparison of MHD shock stand-off enhancement (\% increase in leading shock distance from surface) for the experimental condition with magnetic field strength $B_{ref} = 0.36$ T: experimental result by Wasai et al. \cite{Wasai2010} (as measured from experimental graphs), and computed result from the numerical model of this work (as measured from computed pressure traces along radial lines shown in FIG. \ref{fig:radial_traces}).}
	\label{tab:MHDtable}
	\begin{ruledtabular}
		\begin{tabular}{p{0.3\linewidth} | p{0.3\linewidth} | p{0.3\linewidth}}
			\begin{center} Radial distance \end{center}  &  \begin{center} MHD shock stand-off enhancement: experiments (\%) \end{center}  &  \begin{center} MHD shock stand-off enhancement: simulations (\%) \end{center}  \\
			\hline
			\begin{center} R = 4 mm \end{center}  & \begin{center} 17.9 - 82.1 \end{center} & \begin{center} 24.1 \end{center} \\ 
			\begin{center} R = 8 mm \end{center}  & \begin{center} 7.9 - 17.5 \end{center} & \begin{center} 8.8 \end{center} \\ 
			\begin{center} R = 10 mm \end{center}  & \begin{center} 7.2 - 25.1 \end{center} & \begin{center} 7.8 \end{center} \\
		\end{tabular}
	\end{ruledtabular}
\end{table}

The experimental work resolves a property trace through lines parallel to the z-axis of symmetry at different radial distances (r = 4mm, 8mm and 10mm) along the model to determine shock position more precisely. These brightness intensity property traces along the different radial lines are shown in FIG. \ref{fig:radial_traces}, and the increase in leading shock position due to MHD effects of the imposed magnetic field can be measured from these graphs. The leading shock position is considered to be where the property trace of brightness intensity first rises in value. Since multiple runs are presented in the one graph, the minimum offset and maximum off-set between the no-\textbf{B} and \textbf{B} = 0.36 T cases of all test runs is measured and a \% increase in the shock stand-off distance is presented as the range of values in Table \ref{tab:MHDtable}. 

%Whilst the measured brightness intensity relates most strongly with temperature, the rendering cannot be mapped directly onto any one property from the simulation for a number of reasons: the brightness intensity is measured as a 2D side view of the 3D experiment, which is different to the 2D planar cross section of the simulation. Pressure is chosen as the property for comparison from the simulation results as it clearly shows the shock front position and retains a fairly steady value through the shock layer to the model surface - which offers a similar profile to the brightness intensity rendering. As per the discussed differences, the two-step profile of the numerical trace at R=4mm shows how the trace covers the separation shock (point at which stand-off enhancement is measured) and then the high pressure impingement point at the surface - a cross-sectional profile detail which cannot be captured by the 2D projected side view of the flow from the experimental images.

The important conclusion from the experimental results is that there is a clearly observable, and measurable, MHD enhancement of shock-stand-off distance. From the pressure traces along the radial lines, this observable increase in stand-off distance is presented. In Table \ref{tab:MHDtable}, the measured increase in leading shock stand-off distance from the simulations of this work is compared to the measured increase from the trace taken along the radial lines in the experimental results published by Wasai et al. \cite{Wasai2010}. The results show an increase in shock stand-off distance which falls within the range measured from the experimental results. Therefore the predictions of this work are quantitatively in-line with experimental measurements for matched magnetic field strength. The agreement in shock stand-off enhancement indicates that the model of this work is able to realistically capture the complex MHD flow control effect. This is a key result for the complex flow field: a hypersonic flow with SWBL interaction over the double cone model coupled with magnetic interaction. 

\subsection{MHD effects verification}

Results of the Nagata et al. numerical simulations showed that for a magnetic field strength matched with the experimental conditions, the MHD effects were negligible in the computed steady state flow field solution (see property traces of FIG. \ref{fig:radial_traces}). However, the experiment does in fact reveal a small but observable increase in shock stand-off distance due to MHD flow control. By initialising the magnetic field strength in the numerical simulation to be much higher than that of the experiment, Nagata et al. were able to demonstrate and explore the qualitative effects an imposed magnetic field has on the flow, which agrees with the qualitative effect observed in the experiment.

As such, in this verification section we seek to compare the qualitative effects observed for the introduced strong magnetic field case ($B_{ref} = 1.5$ T) with the effects reported in the previous numerical study. 

\begin{figure*}[ht]
	\centering
	\includegraphics[width=17.0cm]{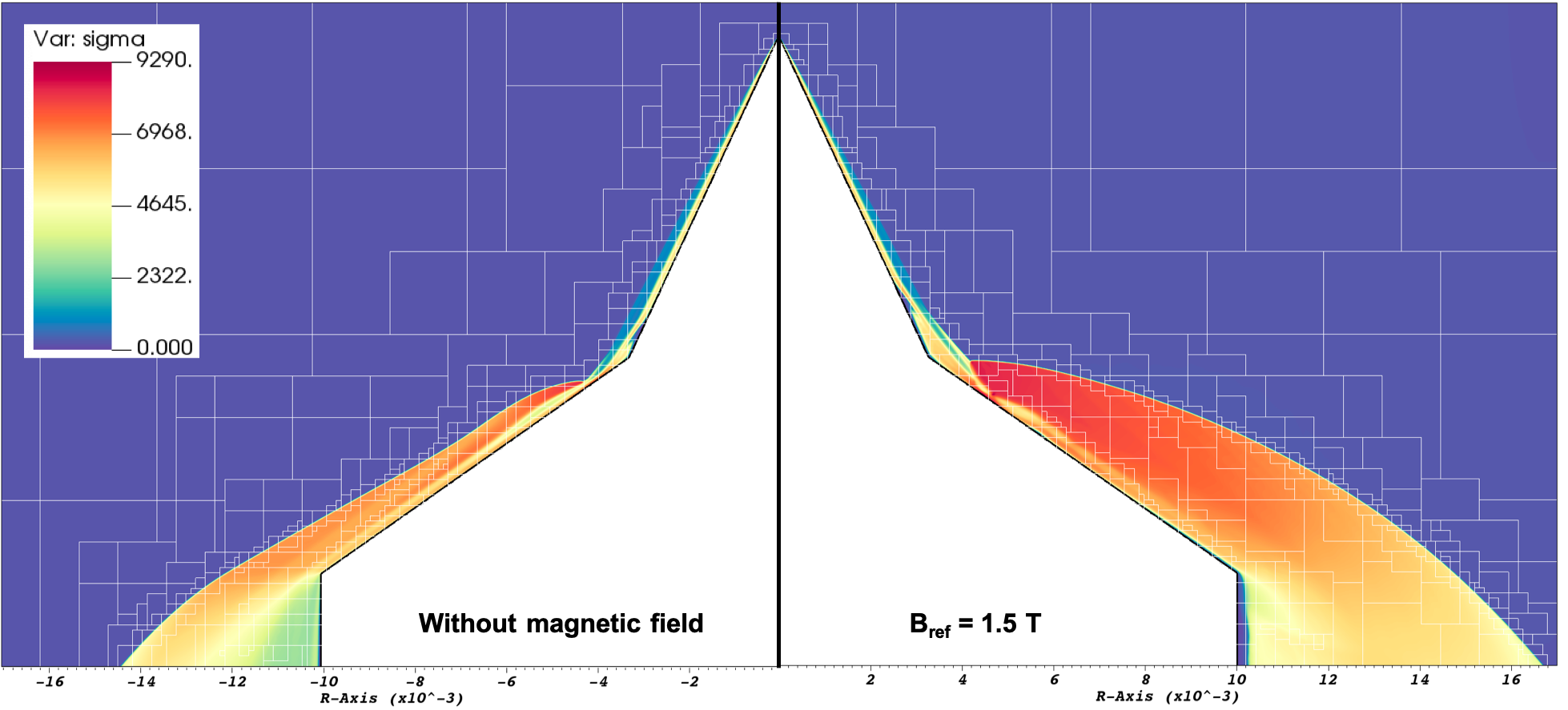}
	\caption{Comparison of steady state solution for electrical conductivity ($\sigma$) without magnetic field (left plane) and for the case of applied magnetic field of strength $B_{ref} = 1.5$ T (right plane). Regions of mesh refinement are indicated by the marked white mesh block borders.} 
	\label{fig:DC_sigma_comparison}
\end{figure*}

A comparison of computed electrical conductivity is also shown for the two cases in FIG. \ref{fig:DC_sigma_comparison}: base flow without magnetic field, and the strong magnetic field case. This plot more clearly shows the qualitative differences between the base flow and an augmented flow under strong magnetic influence. The strong \textbf{B}-field strength result agrees in terms of qualitative phenomenology with the simulation results of previous studies \cite{Nagata2011numerical, Nagata2013Bfield}. Specifically, we observe the MHD effects cause the shock layer over the second plane of the model to lift away from the surface, and for the stronger \textbf{B}-field the shock wave begins to become bow shaped. The separation point moves upstream, and shock triple point becomes extended. The electrical conductivity is a dominant parameter driving the computed Lorentz forcing dynamics, and FIG. \ref{fig:DC_sigma_comparison} shows the enlarged extent of high conductivity plasma in the detached shock region of the $B_{ref} = 1.5$ T test case. 

The Nagata studies were successful in characterising the qualitative effects predicted by strong imposed magnetic fields on the flow augmentation. They concluded in their work that the two key factors limiting the accuracy of the model were: effectively capturing the sensitive feature formation at sufficiently high resolution where shock waves interact, and the accuracy of the thermochemical model (which underpins prediction of electrical conductivity). As such, the model of this work aims to build on these identified factors by utilising high resolution, high efficiency, numerical methods combined with advanced thermochemistry modelling for air plasma and its electrical properties.

%Specifically, it supports the accuracy and suitability of the plasma19X equilibrium-based EoS to compute the thermochemical and electrical properties of the fluid.

\FloatBarrier

\section{Numerical studies of hypersonic flows with imposed MHD effects}
\label{sec:analysis}

\vspace{-3mm}

The validated numerical model is now applied to varied configurations of the hypersonic double cone geometry with an imposed dipole magnetic field. 

Within the motivating context of hypersonic flight control, even small changes in shock position can affect the flight dynamics. For a non-simple shock structure, morphological adaptations - changes of the actual topology of the shock structure - are particularly important to identify and to classify. In relation to the surface actuation concept, the identification of topological changes in shock structure determines whether MHD actuation can morphologically replicate mechanical surface actuation effects. 

These studies seek to explore and explain the effects of magnetic field configurations, combined with variations in geometry, in terms of the quantitative (\% enhancement of detached shock position) and structural adaptations (different shock structure topologies) observed in the emergent flow fields. Through such studies, the conditions and mechanisms of action in the coupled fluid-dynamic and magneto-dynamic flow physics seek to be better understood. Conditions which amplify the MHD enhancement effect, and conditions which produce flows of morphological equivalence between surface geometry and magnetic actuation can be identified. 

\subsection{Shock structure classification system}

\begin{figure}[h!]
	\centering
	\includegraphics[width=6.5cm]{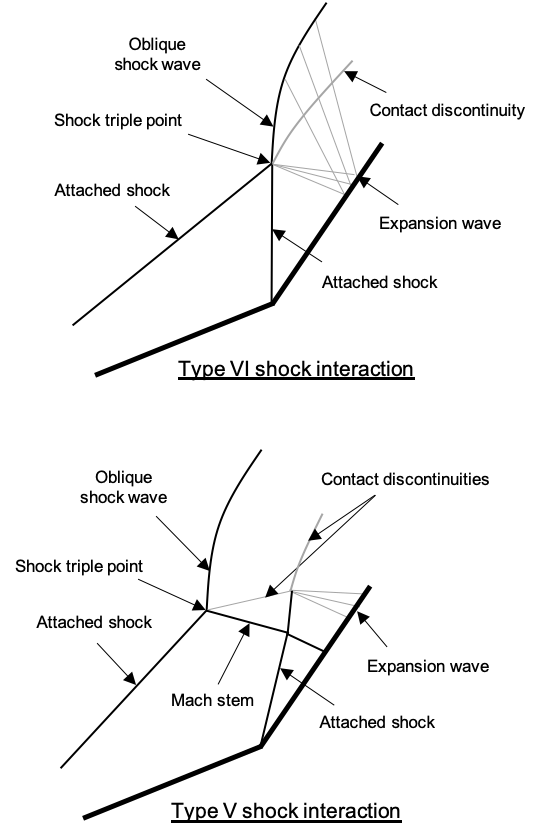}
	\caption{Edney Type VI and Type V classification of shock interaction which occurs locally at the kink point for a 2D inviscid flow over a double wedge.} 
	\label{fig:Contour_Types}
	\vspace{-5mm} 
\end{figure}

The earliest experimental and theoretical work on the classification of shock interactions was conducted by Edney who examined the mechanisms of interaction between an oblique shock wave and a detached bow shock around a cylinder \cite{Edney1968}. The emergent shock interactions were classified as Types I-VI. This work was then extended by Olejniczak et al. to determine inviscid shock interaction classifications when applied to double wedge geometries \cite{Olejniczak1997}. Though the shock structures differ between the cylinder and double wedge cases, corresponding shock interactions for the steady shock structures which form for an inviscid 2D flow over a double wedge can be determined for Types IV, V and VI.  Olejniczak et al. also identify an additional shock interaction Type IVr which occurs over a double wedge but which does not occur in the original cylindrical shock classification system. 

The corresponding Type VI and Type V shock structure classification of the local interaction near the kink for an inviscid flow over a double wedge is depicted in FIG. \ref{fig:Contour_Types}. As can be seen, when neglecting viscous effects the shock interactions involve an attached shock from the kink point of the double wedge geometry. However, this entire region becomes separated in the case of viscous hypersonic flows, therefore forming entirely different types of shock interactions. 

Olejniczak et al. note in a subsequent study that the problem becomes more complex when viscous effects are accounted for \cite{Olejniczak1996}. The viscous SWBL interaction which leads to the formation of a separation region, produces a separation shock which alters the basic structure of the flow as it interacts with the other flow features. Additionally, growth of the separation region can cause the flow to become unsteady under certain freestream and geometric conditions \cite{Swantek2015}. Given the divergent phenomenologies, Olejniczak et al. conclude that it is not always possible, nor appropriate, to relate viscous double wedge flows to Edney's classification system. More recent work examines the complexities of viscous double wedges and control surfaces \cite{Durna2016,Hu2008}, however, flows in this regime do not adhere to existing shock interaction classification systems, and no additional shock classification system is introduced. 

\begin{figure}[h!]
	\centering
	\includegraphics[width=7.3cm]{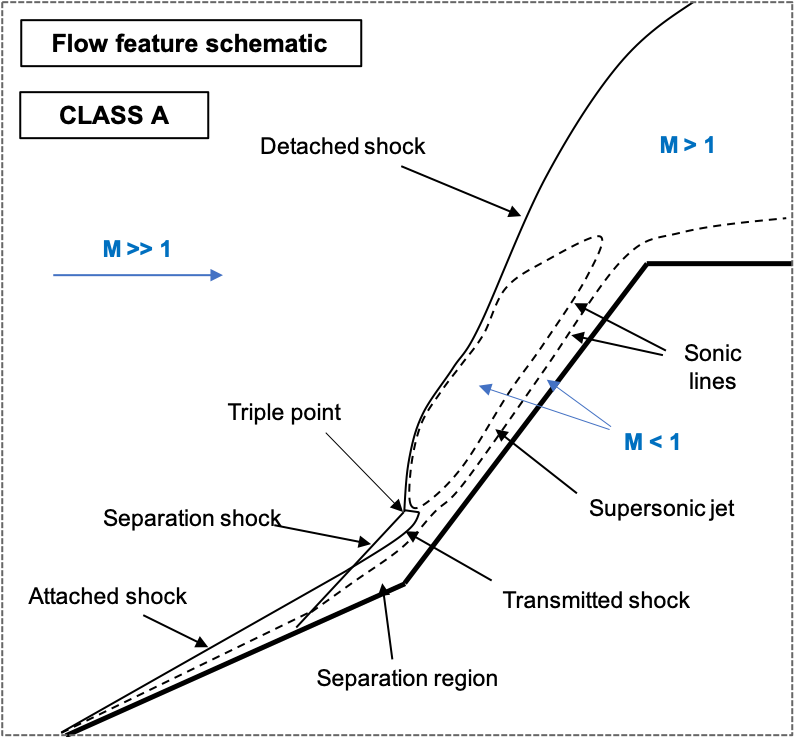}
	\includegraphics[width=7.3cm]{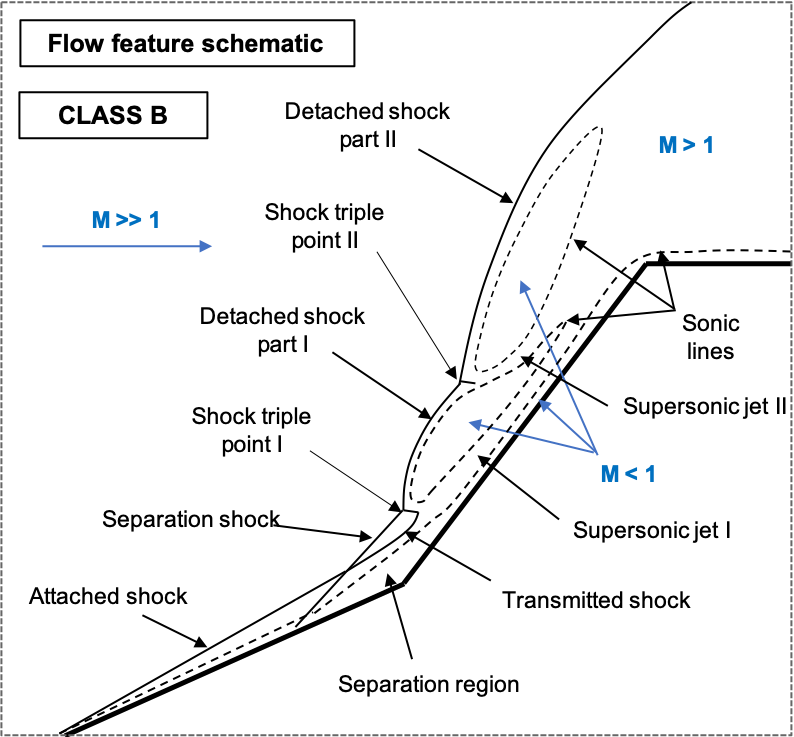}
	\includegraphics[width=7.3cm]{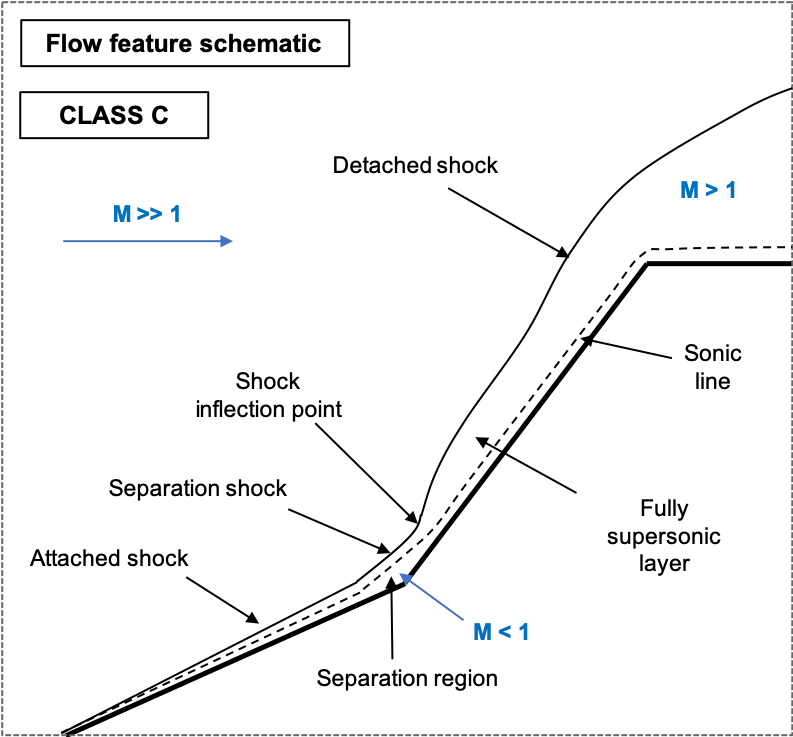}
	\caption{Schematic of the flow structure classification system. All emergent flow fields in this analysis can be identified as one of: CLASS A, CLASS B or CLASS C. The shock features are depicted with solid lines and features of the flow are labelled, including sonic lines (M=1) to depict the supersonic and subsonic regions. } 
	\label{fig:Schematic_Types}
\end{figure}

This work considers viscous axisymmetric hypersonic flows over double cone geometries, as distinct from the double wedge studies. The dimensionality of the cone vs wedge promotes steadiness of the flow. Classification of the emergent steady state flow types serves as an important basis for analysis. Therefore, 3 classifications are proposed for the emergent flow topologies of this work. As distinct from Edney's classification system, and the double wedge equivalences, the classification system is not based on the singular shock interaction arising near the kink point, but the form of the \emph{overall} shock structure as determined by the presence, absence or duplication of salient flow features. Specifically: the formation of shock triple points, and bounded subsonic regions. Therefore the topologies are delineated as classes, which consider \emph{all} shock interactions (multiple may occur) within the total shock structure. 

This classification system is shown in FIG. \ref{fig:Schematic_Types} and will be used in the analysis of this section. CLASS A represents the topology produced by hypersonic double cone tests in the CUBRC validation tests, where a single shock triple point, detached shock, and bounded subsonic region forms. CLASS B identifies flow structures where features surrounding the detached shock region ahead of the second surface become duplicated. Here a second shock triple point is associated with the formation of a second discrete subsonic region and a distinct secondary portion of the detached shock. CLASS C collectively identifies cases where the reflected flow over the second surface produces a shock inflection point, rather than a shock triple point, and no isolated subsonic region forms beyond the boundary layer and separation region. The definitions and constituent flow features of each flow class are represented in the schematics of FIG. \ref{fig:Schematic_Types}. 

\subsection{Variation of second conical surface angle $\theta_2$}

To first examine the effect of varying geometry on the base flow (no magnetic field), we conduct a parameter study of varied second surface angle ($\theta_2$) . Inclination of the second conical surface generically replicates the effect of a mechanical control surface. 

\begin{figure}[h!]
	\centering
	\includegraphics[height=7.2cm]{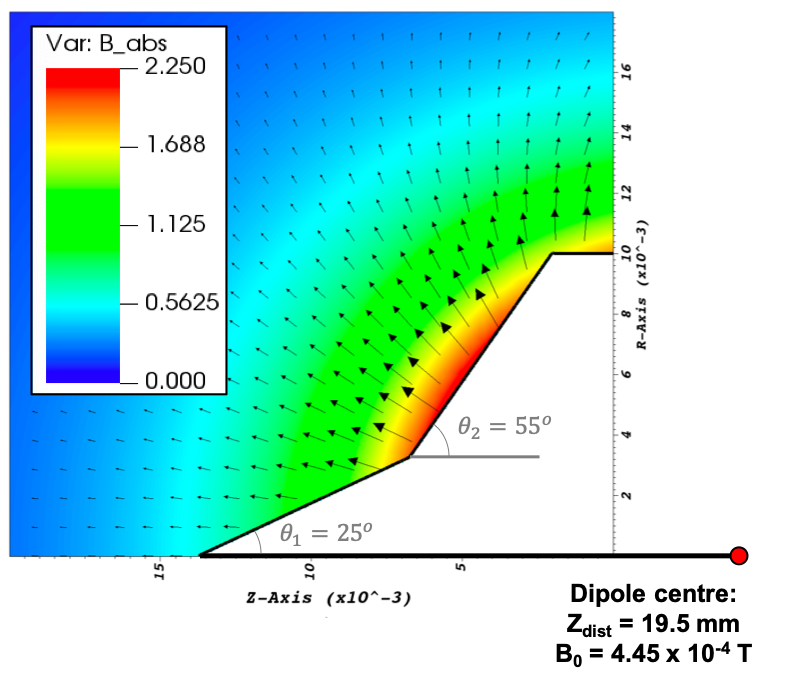}
	\caption{Imposed dipole magnetic field configuration for the $\theta_2 = 55^o$ test case.} 
	\label{fig:B_abs_angle55}
\end{figure}

\begin{figure*}[ht]
	\centering
	\includegraphics[width=18.0cm]{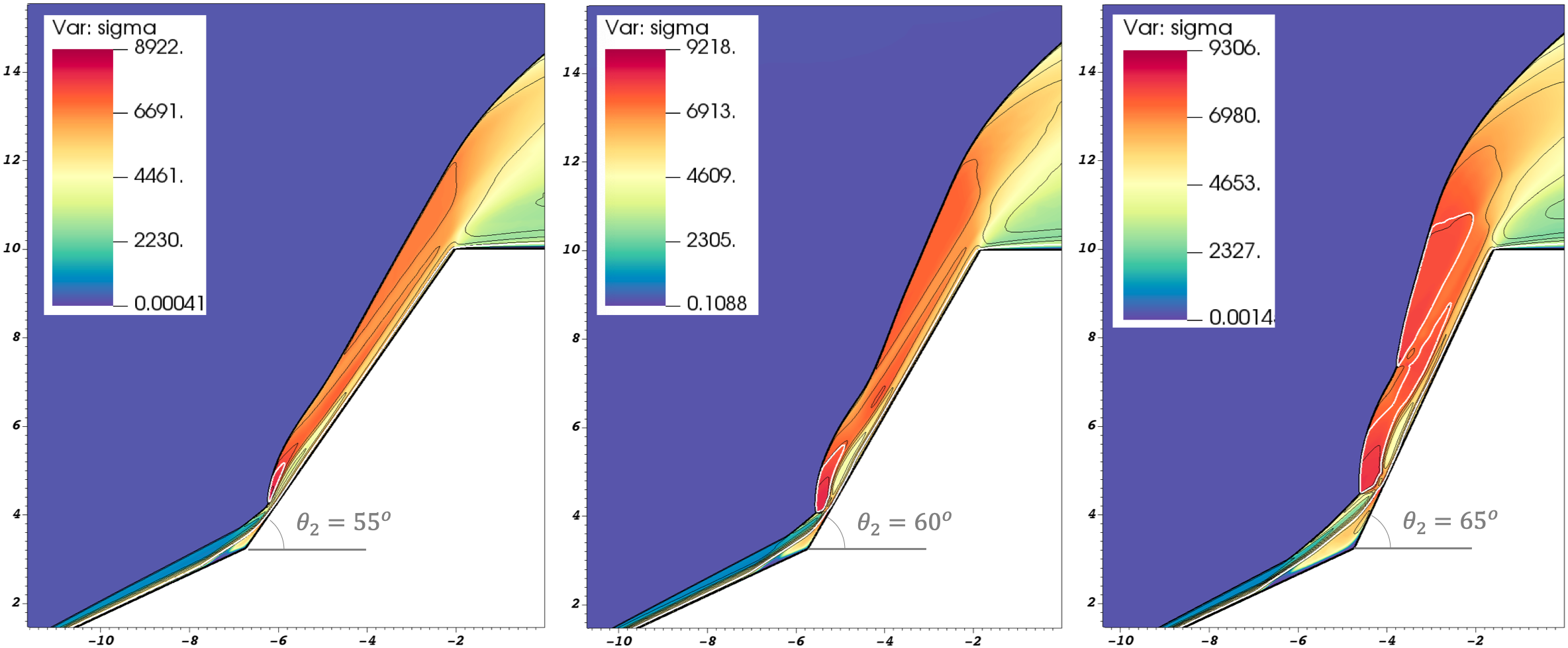}
	\caption{Computed steady state electrical conductivity (colour-map) for the base case without magnetic field activation for each of the second surface angles. Mach number contours are overlaid in black with the M=1.0 sonic line contour additionally overlaid in white. } 
	\label{fig:AP_sigma} 
\end{figure*}

It is identified from the previous studies that magnetic interaction is dominant in regions where the flow is at high temperature (high conductivity) and low velocity (low inertial forcing). Lorentz forcing has a linear dependency on $|u|$ (see equation \ref{eq:LorentzForce}) whereas fluid inertial forces scale with $|u|^2$. Therefore the magnetic interaction parameter: ratio of Lorentz forcing to fluid inertial forces is $Q_{MHD} \propto |u|^{-1}$. Increased flow stagnation and elevated electrical conductivity occur concurrently by increasing the angle of the second conical surface. We therefore conduct a parameter study with $\theta_2$ surface angles of: $55^o$ (experimental condition), $60^o$ and $65^o$, where conditions are expected to \emph{amplify} the shock enhancement effect once a magnetic field is introduced. 

We first examine the resultant steady state base flows for each case: shown in FIG. \ref{fig:AP_sigma}, As anticipated, increasing $\theta_2$ creates larger regions of elevated electrical conductivity, and the formation of larger subsonic flow regions (depicted via the white sonic lines). Most importantly, however, we observe for the $\theta_2=65^o$ inclination case, the emergence of what can be identified as a different shock structure \emph{class} (CLASS B). As per the constituent features of the schematics in FIG. \ref{fig:Schematic_Types}, CLASS B exhibits the emergence of a second shock triple point, which causes the formation of a second subsonic region within the detached shock region over the second surface.

We now introduce the dipole magnetic field in order to study the imposed MHD effects on the flow cases. For consistency with the experimental configuration of Section \ref{sec:MHD_exp_val}, the magnetic field is configured to match the strong \textbf{B}-field case, with the dipole centre located at 19.5 mm from the conical point. The corresponding peak $|B|$ value in the fluid domain is 2.25 T, occurring partially along the second surface, as shown in FIG. \ref{fig:B_abs_angle55} ($B_0$ of equation \ref{eq:B_dipole} to produce this field is cited in figure). 

Varying the second surface angle affects both the maximum |B| in the flow field and the angle of the \textbf{B}-field lines relative to the surface (and flow field). There is no perfect solution to isolate the effect of changed surface angle within the dipole magnetic field, given this is a complexly coupled problem. However, configurations can be achieved which closely control \textbf{B}-field orientation and magnitude relative to the actuation surface. The selected approach is to maintain the dipole centre at $Z_{dist} = 19.5$ mm from the conical point, whilst fixing the shoulder point and kink point of the geometry on the r-axis, and extending along the z-axis as the $\theta_2$ angle changes. This produces a magnetic field which is approximately congruent for each test case angle and $B_0$ is scaled to maintain precisely constant $B_{max}$ of 2.25 T, in the flow field.

%First examining the base case flow results (without magnetic field activation) for each of the surface angles, FIG. \ref{fig:AP_sigma} resolves electrical conductivity with Mach number contours. Results show that increasing $\theta_2$ to $60^o$ elevates the peak electrical conductivity by only $3.32 \%$ as compared to the $\theta_2 = 55^o$ case. The separation region is marginally larger, the stem of the shock triple point becomes extended resulting in the small subsonic region becoming slightly enlarged. Elevation of $\theta_2$ from $60^o - 65^o$ has a more notable effect as it surpasses the angular forcing threshold to cause the formation of a second shock triple point. The increase in peak conductivity is only $4.30 \%$ greater than the $\theta_2 = 55^o$ base flow, however the first subsonic region is already much larger, and a second shock triple point forms downstream, resulting in a second subsonic region. The consequence is that a larger total volume of gas has high electrical conductivity relative to the flow field. This condition is expected to promote MHD enhancement. 

\begin{figure*}[ht]
	\centering
	\includegraphics[width=18.0cm]{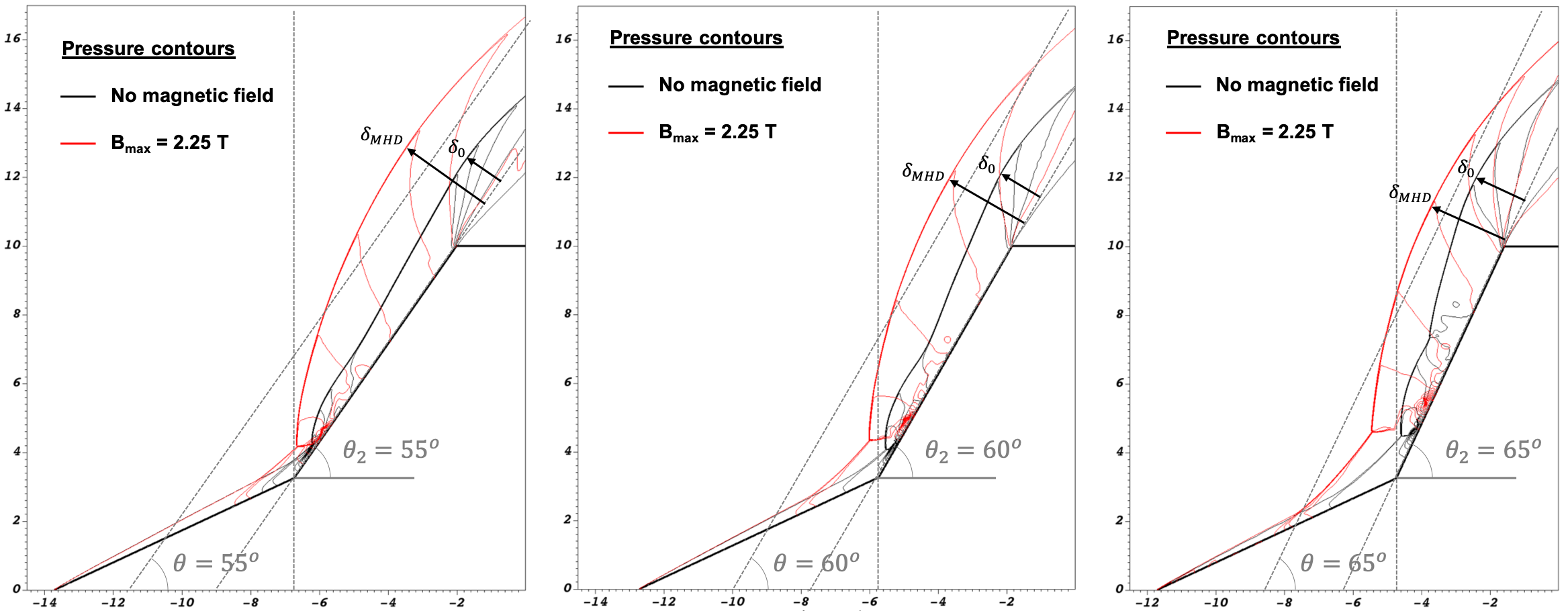}
	\caption{Computed steady state pressure contours for different second conical surface angles. Guide lines are added for reference: one perpendicular to the z-axis of symmetry, intersecting through the kink point, and two tangential guide lines running parallel to the second surface incline, one at the surface and one offset 2 mm from the vehicle surface. } 
	\label{fig:AP_pressure_contours} 
\end{figure*}

FIG. \ref{fig:AP_pressure_contours} shows the steady state pressure contours for the base flow case as well as the enhanced shock structure of the MHD affected case for each conical surface angle. Reference lines are added to aid the analysis of MHD enhancement effect. Two classes of effects are observed as a result of the imposed magnetic field: 
\begin{enumerate}
	\item Alteration of the flow structure classification 
	\item Quantification of the magnitude of shock enhancement effect between cases
\end{enumerate}

Notably, for the $\theta_2 = 65^o$ case, addition of the magnetic field is observed to alter the steady state flow topology from CLASS B to CLASS A. The magnetic enhancement of the separation region alters the fluid dynamics downstream along the second surface, while Lorentz forcing acts to bow the shape of the detached shock in all cases. Combined, the resultant steady state forms to become CLASS A under the imposed MHD effects.
 %The mechanism by which will be examined further. 

%Given the variable morphology of both the base case and MHD affected shock structure, there is no singularly perfect way to quantify MHD enhancement effect. 

With reference to the guide line running tangential to the second surface (off-set at 2mm), for increasing $\theta_2$ values: all 3 inclined surface cases become more bow shaped and show a \emph{decreasing} segmented arc which exceeds the tangential guide line. Measurement of shock stand-off and \% enhancement from the base case is calculated by taking the peak stand-off distance perpendicular to the second surface tangential line for both the base case and MHD affected case. Values are compared in Table \ref{tab:APtable}. 

\begin{table}
	\centering
	\caption{Collated in the table is the measured peak stand-off distances for $B_{max} = 2.25$ T case, with computed MHD stand-off distance calculated as a \% enhancement from the base case: $ enhancement (\%) = \frac{\delta _{MHD} - \delta_0}{\delta_0} \times 100\% $.}
	\label{tab:APtable}
	\begin{ruledtabular}
		\begin{tabular}{p{0.3\linewidth} | p{0.3\linewidth} | p{0.3\linewidth}}
			\begin{center} Second conical surface angle: $\theta_2$ \end{center}  &  \begin{center} Maximum perpendicular stand-off distance: $\delta_{MHD}$ (mm) \end{center}  &  \begin{center} MHD shock stand-off enhancement (\%) \end{center}  \\
			\hline
			\begin{center} $55^o$ \end{center}  & \begin{center} 2.81 \end{center} & \begin{center} 139.15 \end{center} \\ 
			\begin{center} $60^o$ \end{center}  & \begin{center} 2.55 \end{center} & \begin{center} 91.73 \end{center} \\ 
			\begin{center} $65^o$ \end{center}  & \begin{center} 2.38 \end{center} & \begin{center} 61.90 \end{center} \\
		\end{tabular}
	\end{ruledtabular}
\end{table}

\begin{figure*}[ht]
	\centering
	\includegraphics[width=18.0cm]{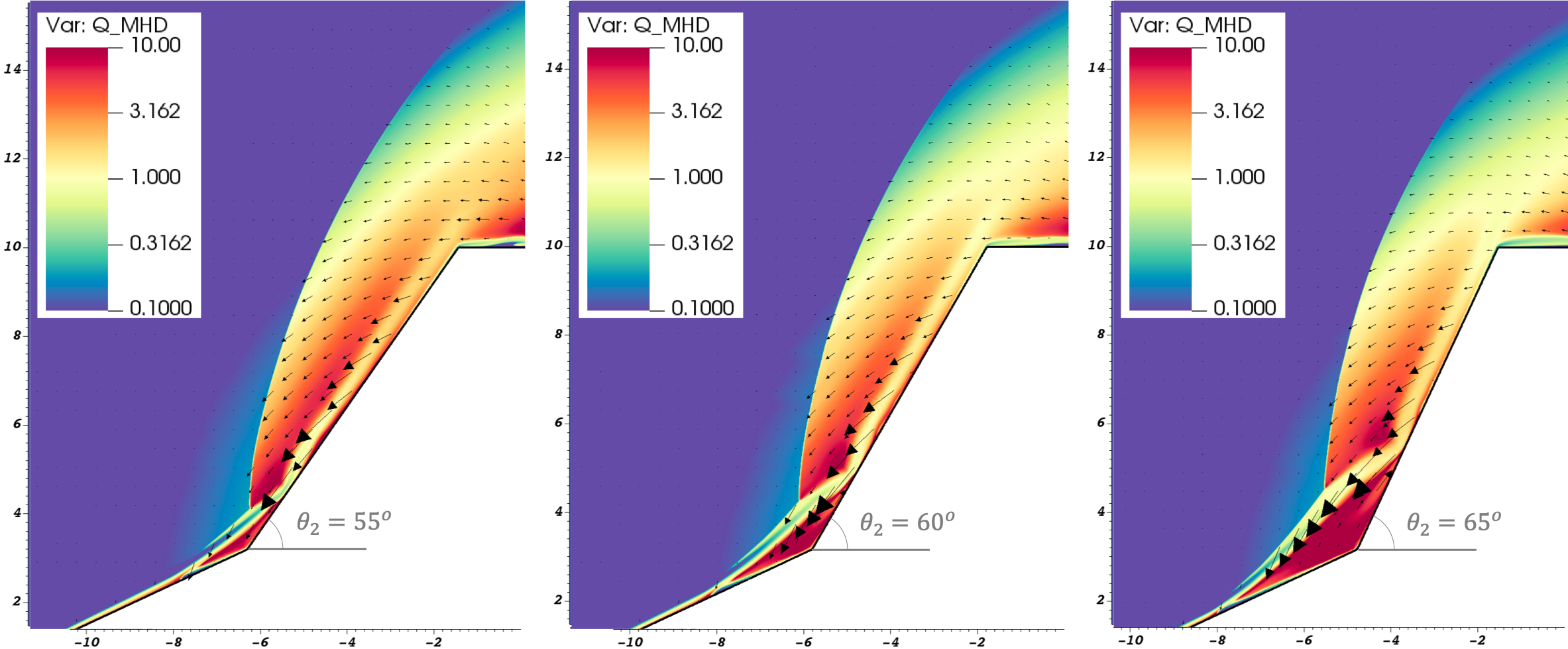}
	\caption{Computed $Q_{MHD}$ and Lorentz forcing vectors for each $\theta_2$ surface angle. } 
	\label{fig:AP_QMHD} 
\end{figure*}

The base flow conditions depicted in FIG. \ref{fig:AP_sigma} indicate increased $\theta_2$ promote conditions which would \emph{amplify} the shock enhancement effect once a magnetic field is introduced. 
The counter-intuitive result which arises, is that the peak stand-off distance and \% enhancement both diminish for increasing $\theta_2$.  The MHD enhanced shock structure lifts maximally away, perpendicularly from the surface lines, for the case of minimum $\theta_2 = 55^o$.  

The separation region, however, becomes significantly more enhanced with increasing surface angle. In the base case flow (without magnetic field activation), the separation region is larger, the more inclined the second surface. The subsequent expansion of the separation region due to magnetic interaction results in maximum enhancement for largest $\theta_2$. 

A plot of magnetic interaction parameter (scaled logarithmically from 0.1-10) is compares the different conical surface angles in Figure \ref{fig:AP_QMHD}. The significant expansion of the separation region can be explained by the strong magnetic interaction which is observed to occur in the high conductivity low velocity separation region near the kink. Expansion of the separation region causes the separation shock to become more detached and inclined relative to the first surface. This causes the jet stream to rise and reflect further along the second surface (peak surface pressure point). Enhancement of the separation region diminishes the volume of fluid within the detached bow shock region where peak Lorentz forcing and magnetic interaction occurs in this section of the flow field. The peak Lorentz forcing vectors shift from the subsonic detached shock region and towards the separation region as the surface angle increases. This coupled effect between the detached shock layer and the separation region leads to smaller maximum stand-off distances in the upper bow shock region for increased $\theta_2$. 

\FloatBarrier

\subsection{Magnetic dipole position}

We now examine the effect of changing the magnetic field configuration on the MHD affected flow. 

\begin{figure*}[ht]
	\centering
	\includegraphics[width=18.0cm]{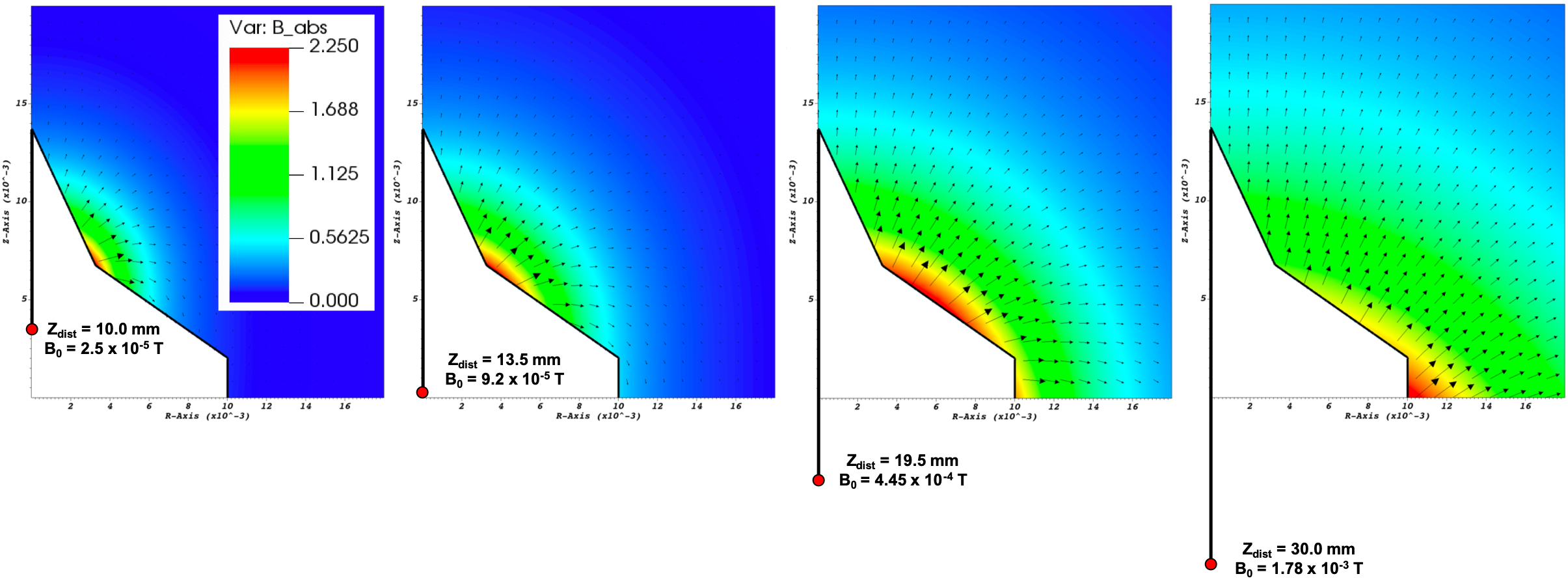}
	\caption{Resultant magnetic fields: intensity colour plot (colour bar shown in left figure holds for all plots) with magnetic field directional vectors scaled to magnitude. Magnetic dipole centre shown, with calculated $B_0$ in order for maximum B-strength in the fluid region to be equivalent at 2.25 T.} 
	\label{fig:DC_Bfields_parameters}
\end{figure*}

Moving the magnetic centre position along the z-axis maintains magnetic axisymmetry whilst altering the orientation of the magnetic field lines in the region where they emanate beyond the vehicle surface. Simply moving the dipole centre location with fixed $B_0$ in the dipole equation means the maximum \textbf{B}-field strength interacting with the fluid also varies. Note that |B| diminishes cubically with radial distance from the dipole centre. To best isolate the effect of \textbf{B}-field \emph{orientation}, independent of \textbf{B}-field \emph{strength}, $B_0$ in the magnetic field equation is scaled such that the $B_{max}$ in the fluid region is maintained at constant 2.25 T. The $B_{max} = 2.25$ T case at $Z_{dist} = 19.5$ mm corresponds to the $B_{ref} = 1.5$ T case of the previous section. FIG \ref{fig:DC_Bfields_parameters} shows how the varied position of dipole centre  changes the orientation of emerging field lines and the location of peak magnetic field strength.

\begin{figure}[ht]
	\centering
	\includegraphics[width=8.0cm]{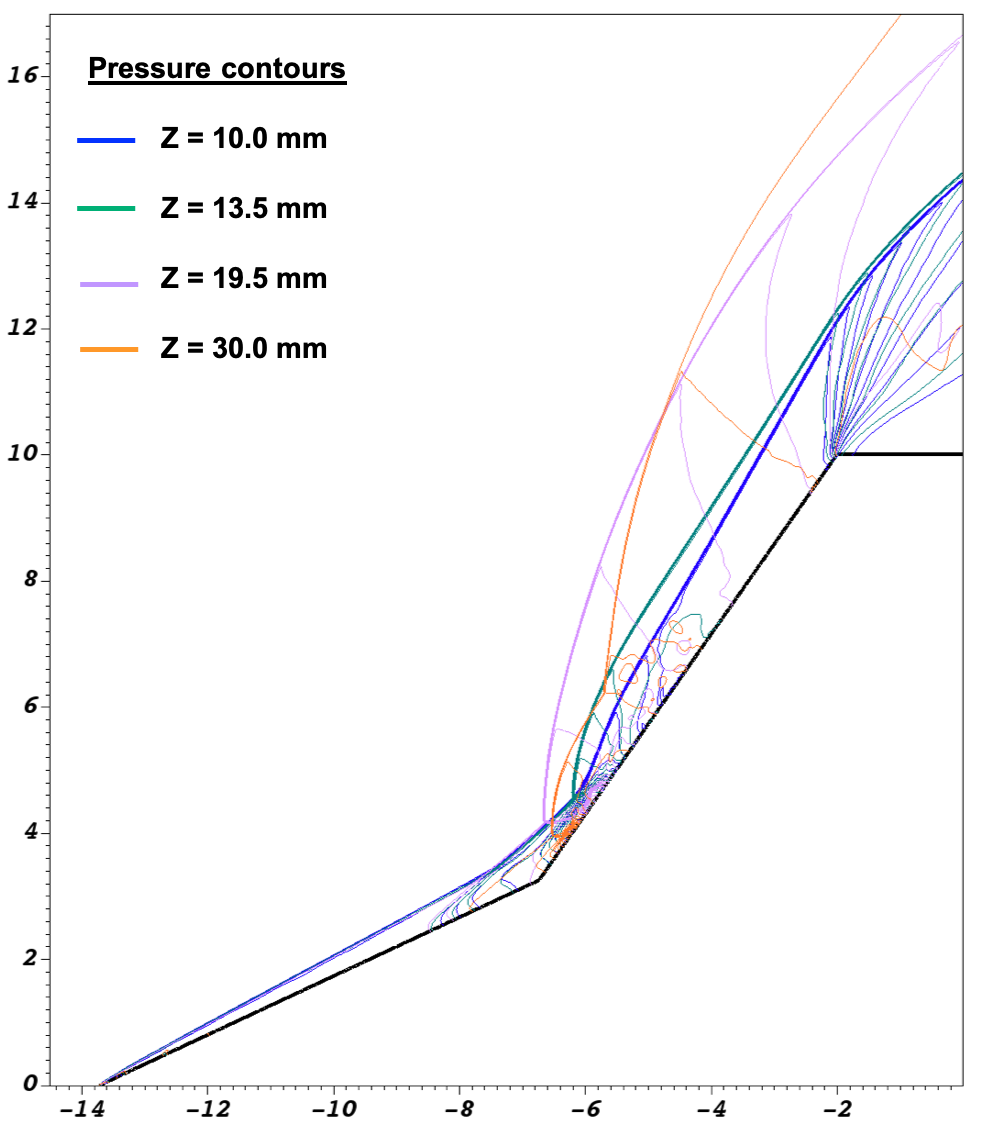}
	\caption{Steady state pressure contours for each magnetic field location. } 
	\label{fig:BP_pressure_contours}
	\vspace{-2mm}
\end{figure}

FIG. \ref{fig:BP_pressure_contours} compares the steady state pressure profiles of the resultant MHD affected steady state flow field for the different dipole centre locations. Several interesting features arise: resultant shock structures exhibit dimorphic overall topologies between cases, which has consequences for the magnitude and characterisation of the shock enhancement effect. 

Most critically, we observe that the $Z_{dist} = 30.0$ mm case leads to the emergence of the CLASS B flow structure (where the non-magnetic base flow was of CLASS A). Further, for the case where the magnetic field is at position $Z_{dist} = 10.0$ mm, we observe the emergence of the CLASS C flow structure. The schematic for the CLASS C flow structure is given in FIG. \ref{fig:Schematic_Types}. As per the classification schematics, the flow is characterised by a shock inflection point, rather than the formation of a triple point, ahead of where the flow impinges on the second surface. This results in a fully supersonic post-shock layer within the detached shock region. 

%\begin{figure}[h!]
%	\centering
%	\includegraphics[height=8.0cm]{./vis/Schematic_TypeC.png}
%	\caption{Schematic of the flow structure classified as TYPE C. Flow features are labelled, including the sonic line (M=1) to depict the subsonic separation region and boundary layer. } 
%	\label{fig:Schematic_TypeC}
%\end{figure}

Examining the general shape of the shock structures of FIG \ref{fig:BP_pressure_contours}, the leading shock front over the second surface lifts away from the vehicle becoming broader above the shoulder for increasing $Z_{dist}$, until the $Z_{dist} = 30.0$ mm case, where we observe the emergence of the CLASS B flow structure. Since the shock wave splits midway along the second surface to form a second shock triple point, in the region centred at approximately the 6 mm radial line, the distance of the leading shock from the vehicle surface does not behave predictably with magnetic dipole position. Of the tested \textbf{B}-field strength and dipole positions, the $Z_{dist} = 19.5$ mm case creates the largest shock enhancement effect over the region directly forward (z-direction) of the second surface. The detached shock part II of the $Z_{dist} = 30.0$ mm becomes more prominently enhanced in the region above the shoulder. 

Also, notably, the emergent CLASS C flow structure of the $Z_{dist} = 10.0$ mm case results in a more suppressed shock structure. Compared to the base flow case, the detached shock remains closer to the second conical surface. Therefore, this magnetic field configuration could be considered to exhibit an MHD \emph{diminishing} effect. 

\begin{figure*}[ht]
	\centering
	\includegraphics[width=7.5cm]{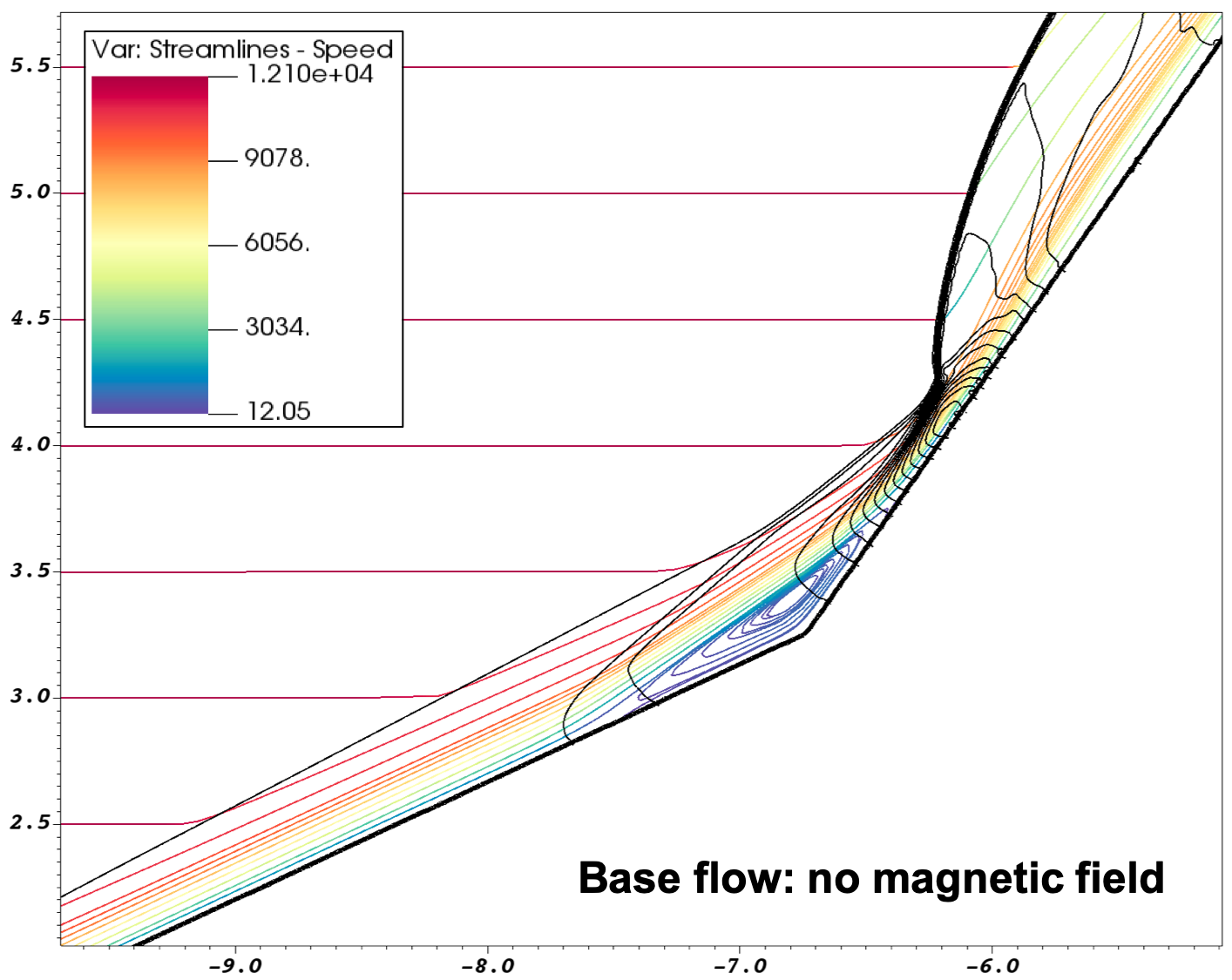}
	\includegraphics[width=15.0cm]{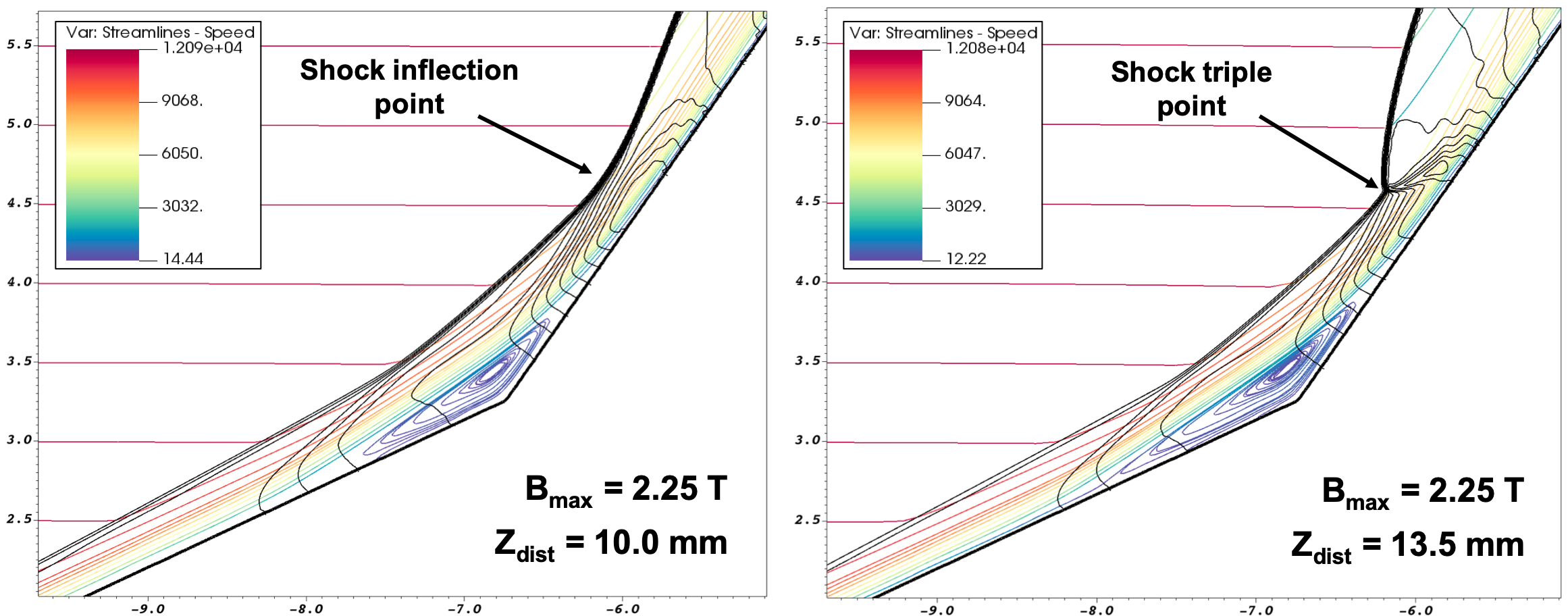}
	\includegraphics[width=15.0cm]{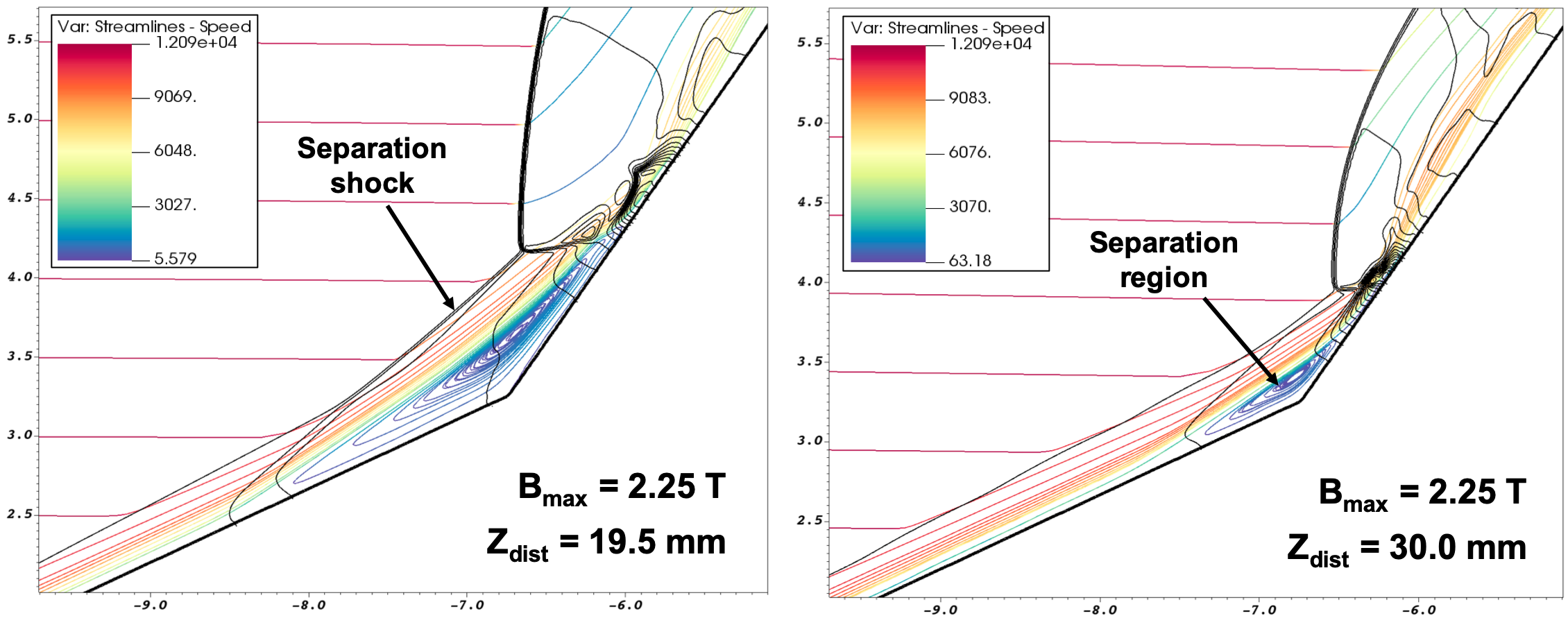}
	\caption{Examining the effect of magnetic field location on separation region: steady state pressure contours in black with computed velocity streamlines (coloured by total velocity magnitude) shown for each magnetic dipole centre location.} 
	\label{fig:BP_streamlines}
	\vspace{-2mm}
\end{figure*}

Focussing our attention on the separation region and surrounds, the position of the first triple point or shock reflection point (depending on the case) is not affected in a predictable way as the peak |B| location moves down the second surface. This is due to the complex interaction of the MHD effect, the SWBLI and separation region formation. 

\begin{figure*}[ht!]
	\centering
	\includegraphics[width=14.0cm]{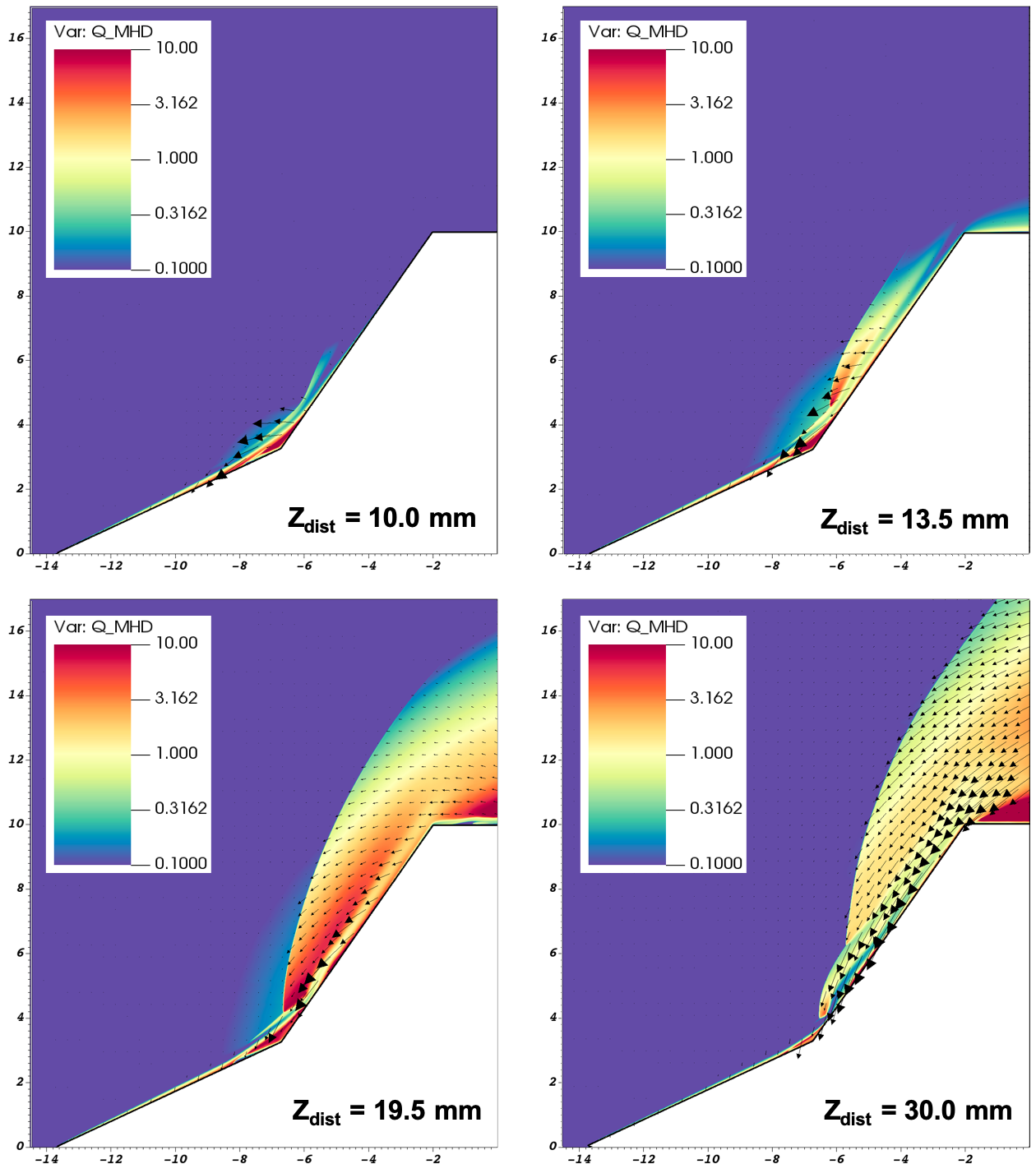}
	\caption{Computed $Q_{MHD}$ (logarithmically scaled between 0.1-10) and Lorentz forcing vectors (showing direction and relative magnitude) for each magnetic field location.} 
	\label{fig:BP_all}
	\vspace{-2mm}
\end{figure*}

Analysis of the separation region (FIG. \ref{fig:BP_streamlines}) in conjunction with the magnetic interaction parameter and Lorentz forcing directional vectors (FIG. \ref{fig:BP_all} characterises some of the interdependent physical phenomena. With reference to these figures, an examination of the flow physics is detailed for each flow structure classification.

\vspace{3mm}

\underline{CLASS C flow structure: $Z_{dist} = 10.0$ mm case} \\ 

As shown in the $Q_{MHD}$ with Lorentz forcing vectors plot, the peak forcing and magnetic interaction both occur close to the kink point. MHD forced deceleration of the flow at the kink causes the separation region to enlarge (see leading pressure contours of the separation region have migrated upstream along the first surface) which causes the flow to be deflected away from the kink at an earlier separation point, resulting in a separated shock which becomes reflected by the second surface further downstream than the base case. Since the flow behind the separated shock is more tangentially aligned with the incline of the second surface (as compared with the base flow), the shock is weakly reflected without forming a shock triple point. Absence of the triple point also eliminates the small stagnation region which forms just downstream of this shock feature (stagnation region retained in all the other flow cases). Note that the colour-mapped streamlines reveal a much higher (and supersonic) velocity magnitude behind the shock reflection point than the shock triple point of the other test cases. Greater velocity magnitude in this region diminishes $Q_{MHD}$ via two concurrent mechanisms: the non-stagnated gas results in a lower static temperature and thereby lower resultant electrical conductivity, while simultaneously the fluid maintains greater inertial force. These aspects dually act to diminish magnetic interaction. Recall that: $Q_{MHD} = (\sigma(p,T) |B|^2 R_0)/(\rho |\textbf{u}|)$, and the flow exhibits both diminished $\sigma(T) $ and increased $|\textbf{u}|$. Consequently, as there is almost no magnetic interaction downstream of this reflection point, there is no MHD forced deceleration to bow the shock wave, and the shock position retains a flattened profile along the second surface. Therefore, counter-intuitively, the imposed \textbf{B}-field actually serves to diminish the overall width of the shock layer over the second surface. 

To summarise, the CLASS C structure forms primarily since the peak \textbf{B}-field acts close to the kink point, enhancing the separation region without causing additional flow deceleration within the detached shock region. Combined, this results in a weak shock reflection instead of a shock triple point which flattens the downstream shock structure. \par 

\vspace{3mm}

\underline{CLASS A flow structure: $Z_{dist} = 13.5$ and $19.5$ mm cases} \\ 

In the $Z_{dist} = 13.5$ mm test cases, the peak magnetic field value is still located very close to the kink, however, in the previous $Z_{dist} = 10.0$ mm case the magnetic field lines were almost perpendicular to the z-axis at the kink, whereas in this dipole position the magnetic field lines emanate at approximately $40^o$ away from the z-axis at the kink point, and the magnetic field diminishes more gradually at this radial distance from the dipole centre (refer to FIG. \ref{fig:DC_Bfields_parameters}). As a result, the Lorentz forcing vectors are also rotated. They now align to be almost direct deceleration vectors near the separation region (vectors run parallel to streamlines, forcing in the opposing direction to the fluid velocity), and the separation region becomes more enhanced in this test case. FIG. \ref{fig:BP_streamlines} shows the vortical streamlines of the separation region are notably enhanced from the base case flow (high $Q_{MHD}$ can be seen in this region of high conductivity and low momentum) which in turn makes the separation shock more detached and lifts the flow reflection point further along the inclined second surface. For this magnetic field position significantly greater Lorentz forcing occurs along the second surface. Lorentz forcing augments the flow further outwards from the second surface resulting in the formation of a shock triple point and subsonic downstream region which further enhances magnetic interaction due to the low flow momentum and high conductivity.  As a result we see a notably bowed shock structure lifted away from the second surface, predominantly in the region just downstream of the shock triple point. The position of the shock wave above the shoulder eventually converges closely with the $Z_{dist} = 10.0$ mm case, where magnetic interaction is weak or absent. \par 

\begin{figure*}[ht]
	\centering
	\includegraphics[width=17.0cm]{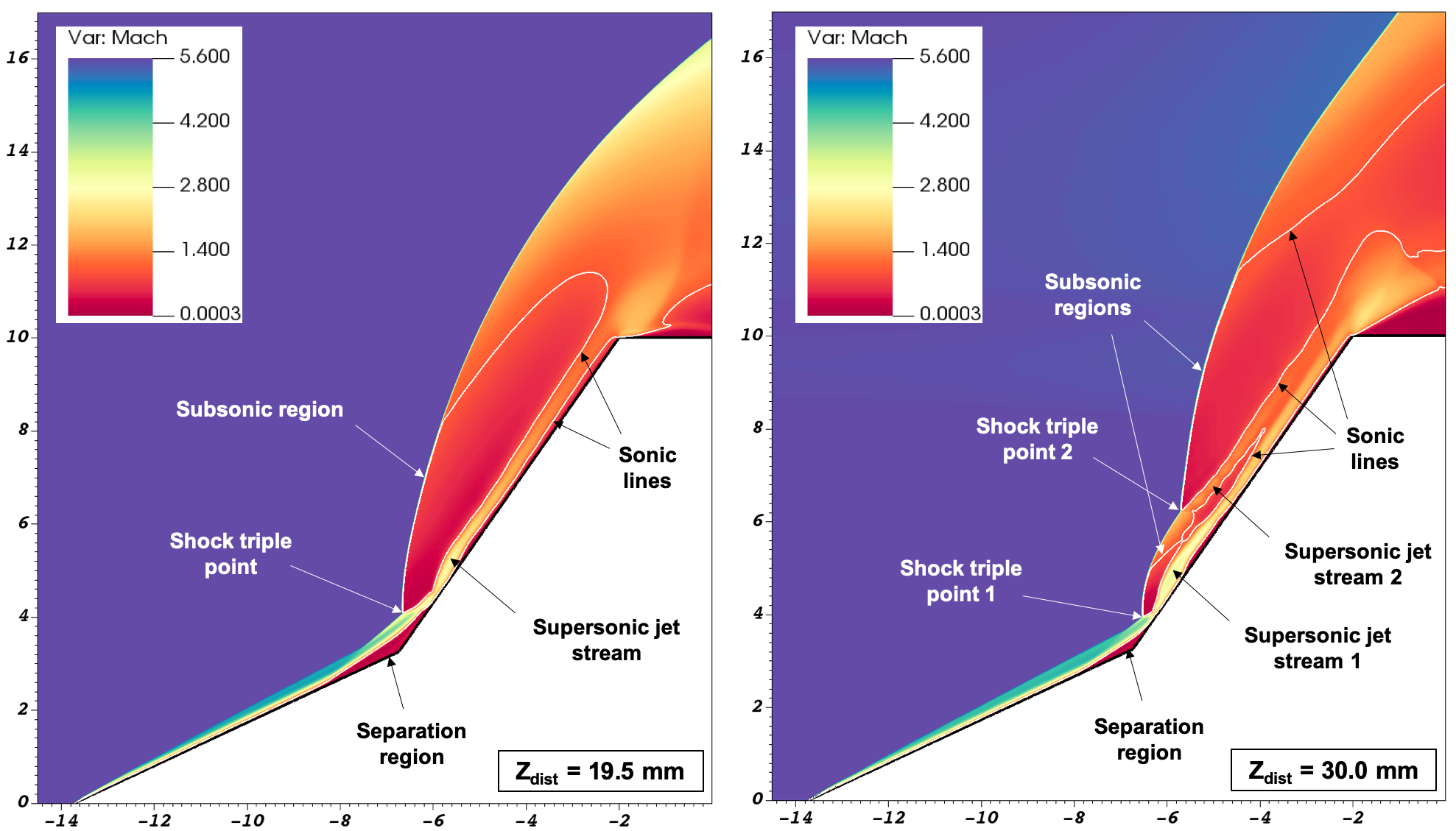}
	\caption{Mach number plots of $Z_{dist} = 19.5$ mm and $Z_{dist} = 30.0$ mm cases, with sonic line ($M=1$) contour and key features labelled.} 
	\label{fig:BP_Mach_plots} 
	\vspace{-2mm}
\end{figure*}

The $Z_{dist} = 19.5$ mm case exhibits the same primary dynamics. However, with peak |B| shifted further down the second surface, the magnetic interaction and Lorentz forcing is more significant in the detached shock region. This results in a more bowed shape of the detached shock and prolonged stem of the shock triple point. This creates the coupled effect of increasing the size of the subsonic high-conductivity region, which amplifies the total shock enhancement effect within the detached shock layer. 

\vspace{3mm}

\underline{CLASS B flow structure: $Z_{dist} = 30.0$ mm case} \\ 

By placing the magnetic dipole centre far from the leading conical point, the peak |B| emerges just above the vehicle shoulder, and the magnetic field diminishes more gradually in intensity. The region of the dipole magnetic field that interacts with the flow, exhibits field lines which are more homogeneously oriented (less variation in the angles of the \textbf{B}-field vectors across the domain). The resultant flow field is very interesting for this test case. The Lorentz forcing acts to directly decelerate the flow velocity over the vast majority of the second surface shock layer (forcing vectors shown to almost directly oppose flow motion). The strong MHD deceleration forces which persist in the region around and above the shoulder effectively creates torque on the leading shock structure, which accumulates in angular force as the shock stand-off distance increases above the shoulder (torque is larger for forcing at a greater displacement), ultimately leading to a split in the detached shock and a second shock triple point being generated. This significantly alters the resultant shock structure (formation of CLASS B from CLASS A base flow), as seen through the comparison of pressure contours (FIG. \ref{fig:BP_pressure_contours}). 

A Mach number plot of the $Z_{dist} = 19.5$ mm and $Z_{dist} = 30.0$ mm cases are compared in FIG. \ref{fig:BP_Mach_plots}. The split shock structure has a number of consequences: a second supersonic jet stream forms in the second surface shock layer, as well as a large second subsonic region forming downstream of the second shock triple point. The consequence of the shock splitting is that MHD enhancement effect is diminished (in terms of \% increase in shock stand-off distance) along the first portion of the second surface, but remains significant above the shoulder. Precise prediction of magnetic dipole position and \textbf{B}-field strength combinations which reach the angular force threshold for shock splitting would be of high importance for design in MHD flow control. 

\vspace{3mm}

\underline{All cases} \\ 

In terms of maximising the MHD enhancement effect for a given magnetic field strength, the orientation of the magnetic field lines are best optimised when oriented perpendicularly to the fluid flow direction, since:
%\vspace{-3mm}
\begin{equation}
\text{Lorentz Force} = \sigma \cdot (\textbf{u} \times \textbf{B}) \times \textbf{B}
\label{eq:LorentzForce}
\end{equation}

where $\sigma$ is a scalar. The $ (\textbf{u} \times \textbf{B}) = |u| |B| \cdot \sin(\theta)$ term is maximised relative to $u$ and $B$ magnitudes when $\theta = 90^o$. There is no singularly clear \textbf{B}-field orientation, however, to achieve this. Since the dipole configuration of \textbf{B} has field lines which vary gradually in orientation, of more significant consideration is that the complex flow field exhibits velocities in many directions at different electrical conductivities. Moreover, the interaction between the flow-field and the MHD forcing affects the cumulative orientation of conductive flow velocity with the magnetic field lines, and so the base flow (without magnetic field) cannot entirely guide predictions of the resultant magnetically influenced flow. These coupled effects, among others, mean that the magnetically affected flow field can behave in counter-intuitive and unpredictable ways - as demonstrated through the cases simulated in this study.

\FloatBarrier

\section{Control surface replacement - an outlook}

In light of the complexity of the flow field coupled with MHD forcing effects, design for precise equivalence between magnetic interaction and mechanical surface actuation is a precarious task. However, an important result which can be drawn from these studies, is the identification of \emph{classes} of configurations which produce \emph{morphology equivalence} in terms of the flow structure classification. 

Since the conditions leading the CLASS A flow structure have been extensively examined, here we demonstrate the two identified classes of cases where CLASS C and CLASS B flow structures emerge: drawing equivalence between surface inclination and magnetic field activation. 

\begin{figure}[h!]
	\centering
	\includegraphics[height=7.5cm]{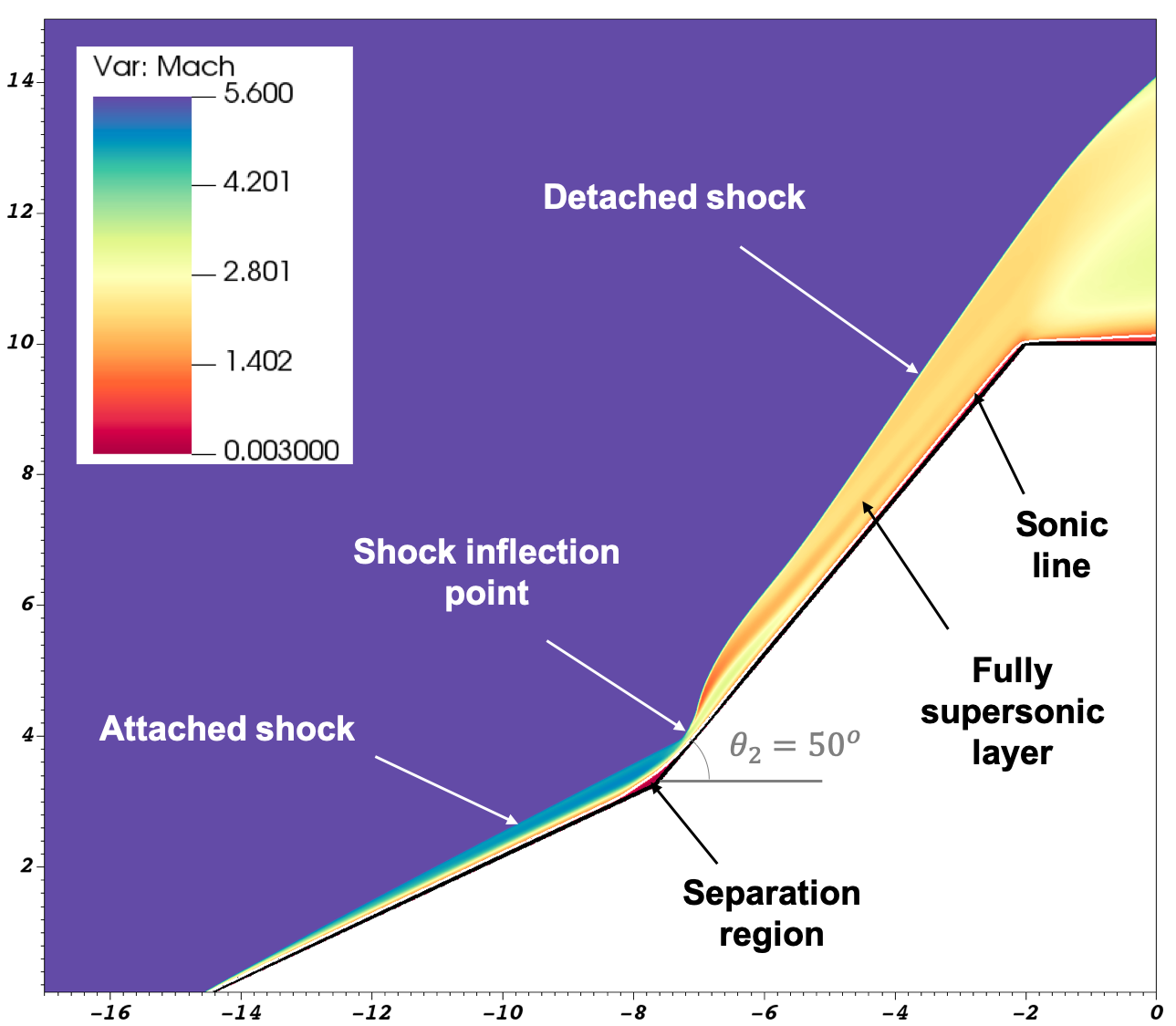}
	\caption{Plot of Mach number for the base flow case with declined second surface angle at $\theta_2 = 50^o$. The steady state flow is of CLASS C exhibiting a shock inflection point and fully supersonic detached shock layer. } 
	\label{fig:Base_angle50}
\end{figure}

Firstly, we introduce the additional case where the second surface angle is \emph{declined}. The base flow case (no magnetic field) is depicted in FIG. \ref{fig:Base_angle50}, and demonstrates the formation of a CLASS C flow structure: a sonic line (M=1) contour is overlaid, demonstrating that only the boundary layer and separation region are subsonic. The remaining detached shock layer is fully supersonic, downstream of the shock inflection point.  

\begin{figure*}[ht]
	\centering
	\includegraphics[width=17.0cm]{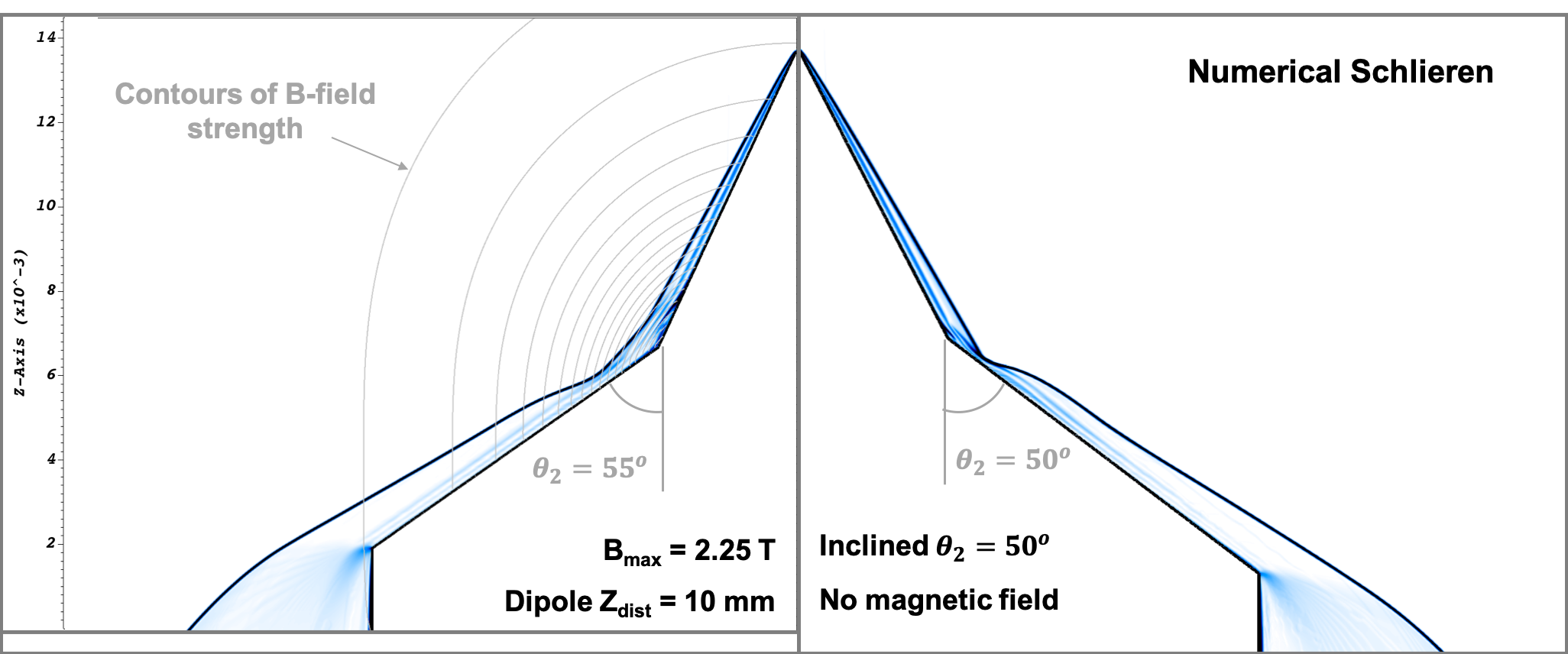}\\
	\vspace{4mm}
	\includegraphics[width=17.0cm]{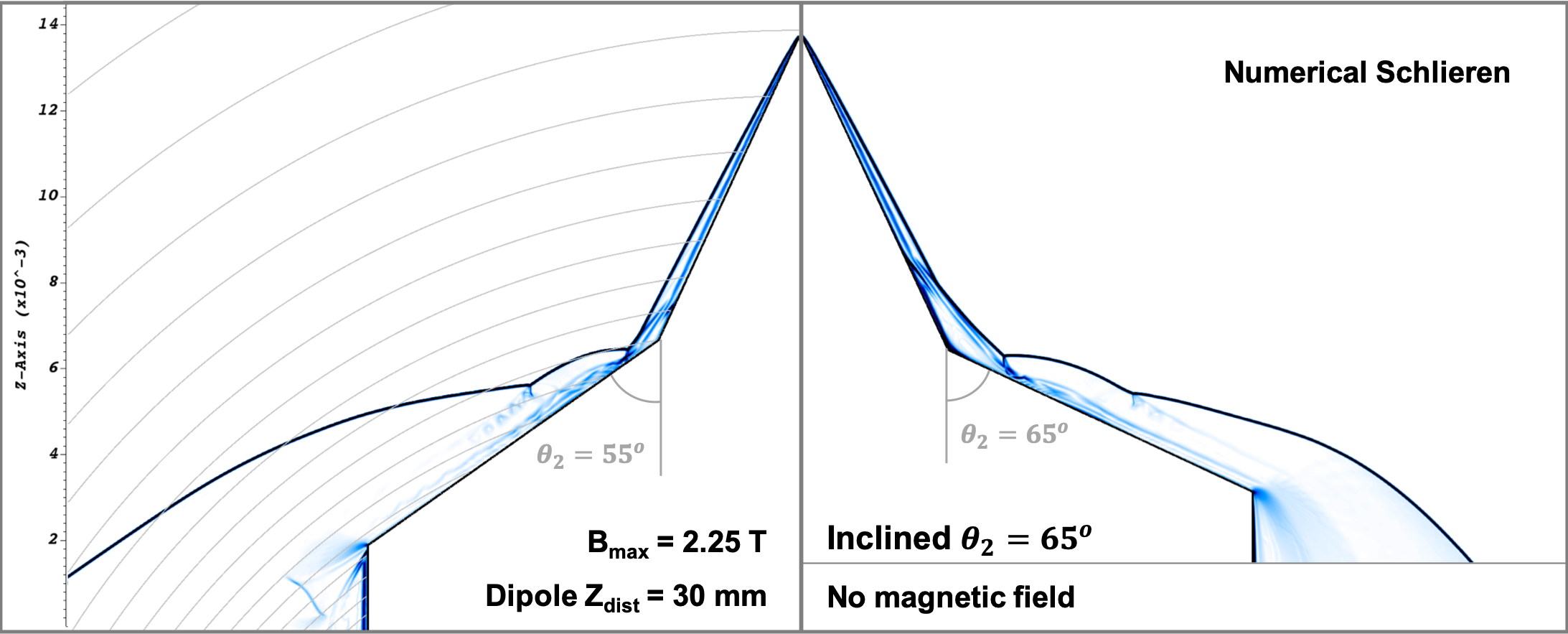}
	\caption{Numerical Schlieren (density gradient) depicting two classes of emergent flow structures: top) CLASS C ,and bottom) CLASS B. In each case an equivalence is drawn between: left) magnetic actuation effect, with contours of |B| depicted in grey, and right) mechanical surface actuation via varied $\theta_2$ angles of the geometry.} 
	\label{fig:Schlieren_TYPES} 
\end{figure*}

The numerical Schlieren images shown in FIG. \ref{fig:Schlieren_TYPES} therefore show two classes of equivalence. The declined second surface angle of $\theta_2 = 50^o$ and the magnetic dipole configuration of $Z_{dist} = 10$ mm produce flow topologies which can both be identified as CLASS C. The magnetic field configuration could be considered to exhibit an MHD \emph{diminishing effect}, with the same topology of a declined actuation surface which similarly produces a thin detached shock layer.

The second Schlieren compares the computed flow field for the $B_{max} = 2.25$ T, $Z_{dist} = 30$ mm case from the dipole location parameter study, with the base flow (no magnetic field) of the $\theta_2 = 65^o$ case. As could be expected, both actions (magnetic and geometric) serve to augment the detached shock position further upstream. The particularly interesting result, is that both conditions produce a morphological adaptation, resulting in the CLASS B flow structure. In both cases, the flow deceleration induces the formation of a second triple point (with second jet stream exhibiting a cascade of internal reflected shocks - seen weakly in the numerical Schlieren). 

Whilst the same class of flow topology can be identified between cases, finer internal differences remain. The size of the separation regions and extent of the detached shock above the vehicle corner, differ between cases. Perfect matching would not be expected given the notable differences in the composition of forcing dynamics. However, identification of equivalent classes of structures, is very important, since it provides a basis for quantitative analysis (regarding enhancement effects), which otherwise cannot be translated directly between flow structure classes. 

%\FloatBarrier

\section{Conclusion}

%MHD affected flows of this type offer potential for the new technology concept - replacement of mechanical surface actuation for flow control. The double cone is a particularly suitable geometry as it produces the complex shock and SWBL interactions which are characteristic of control surface actuation.

Numerical simulation capabilities have been developed for MHD affected, hypersonic flows over 2D axisymmetric non-simple geometries. Producing quantitatively accurate results for this flow type is a challenge due to the sensitivity of feature formation and real gas modelling of the weakly ionised plasma, including the accurate computation of electrical transport properties. The accurate results (as matched with experiment) presented in this work can be attributed to the combination of: effective high resolution numerical methods for strong shock-boundary interactions over arbitrarily complex embedded geometries, and the direct and accurate prediction of electrical conductivity via the extended 19 species air plasma EoS (plasma19X).

Application of the numerical model to studies of the hypersonic double cone configuration, with applied magnetic field, reveal detailed - and at times counter-intuitive - flow field effects. 

A classification system is introduced to aid the analysis of emergent flow topologies. The numerical study of magnetic dipole centre location finds that the flow field is sensitive to magnetic field configuration and geometric variations. The study identifies two classes of effects: (1) differences in magnitudes of MHD enhancement effect, and (2) alteration of the flow topology. One such counter-intuitive result is the reduced MHD enhancement effect for increased second conical surface angle. The detailed flow physics for each topological flow structure is described, revealing complex and coupled mechanisms of action. 

Critically, this paper demonstrates how, for hypersonic flows with complex shock interactions, the MHD affected flow is not only augmented in terms of shock structure position, but may exhibit topological adaptations to the flow structure, as compared with the base case.

In terms of the surface actuation concept, classes of morphological equivalence are identified between the magnetic interaction effect and mechanical control surface actuation. This is of technological significance for emerging magnetic actuation technologies.

\section*{Acknowledgements}
The Authors would like to thank and acknowledge Leo Christodoulou for the number of helpful conversations about the content of this paper. Acknowledgement also to The General Sir John Monash Foundation for being the primary funder of this research, and to the Cambridge Laboratory for Scientific Computing for additional funding.

%\nocite{*}

\FloatBarrier 
%\pagebreak

\section*{List of references}
\vspace{-5mm}
\bibliography{references}% Produces the bibliography via BibTeX.

\end{document}